# *The Illusory Normativity of Rights-Based AI Regulation*

Yiyang Mei & Matthew Sag**

## Abstract

Whether and how to regulate AI is now a central question of governance. Across academic, policy, and international legal circles, the European Union is widely treated as the normative leader in this space. Its regulatory framework, anchored in the General Data Protection Regulation, the Digital Services and Markets Acts, and the AI Act, is often portrayed as a principled model grounded in fundamental rights. This Article challenges that assumption. We argue that the rights-based narrative surrounding EU AI regulation mischaracterizes the logic of its institutional design. While rights language pervades EU legal instruments, its function is managerial, not foundational. These rights operate as tools of administrative ordering, used to mitigate technological disruption, manage geopolitical risk, and preserve systemic balance, rather than as expressions of moral autonomy or democratic consent. Drawing on comparative institutional analysis, we situate EU AI governance within a longer tradition of legal ordering shaped by the need to coordinate power across fragmented jurisdictions. We contrast this approach with the American model, which reflects a different regulatory logic rooted in decentralized authority, sectoral pluralism, and a constitutional preference for innovation and individual autonomy. Through case studies in five key domains—data privacy, cybersecurity, healthcare, labor, and disinformation—we show that EU regulation is not meaningfully rights-driven, as is often claimed. It is instead structured around the containment of institutional risk. Our aim is not to endorse the American model but to reject the presumption that the EU approach reflects a normative ideal that other nations should uncritically adopt. The EU model is best understood as a historically contingent response to its own political conditions, not a template for others to blindly follow.

** Alphabetical order and equal contribution. Yiyang Mei, JD, MPH, is an SJD Candidate at Emory University School of Law (yiyang.mei@emory.edu). Matthew Sag is the Jonas Robitscher Professor of Law in Artificial Intelligence, Machine Learning, and Data Science, Emory University School of Law (msag@emory.edu). Our thanks to George Biashvili, Dave Fagundes, Brett Frischmann, Tonja Jacobi, Margot Kaminski, Matthew Lawrence, Kay Levine, Jonathon Nash, Blake Reid, Jessica Roberts, Andrew Selbst, and participants at the 2025 Legal Scholars Roundtable on Artificial Intelligence and the Emory/UGA Summer Workshop for their comments and suggestions.

# Contents

## INTRODUCTION

Across both sides of the Atlantic, AI regulation has become a proxy for deeper disputes over governance, legitimacy, and the institutional ordering of technological power.[1] The European Union has taken a leading role in global regulatory debates, due in no small part to the perception that its approach is animated by a principled commitment to fundamental rights.[2] Professor Anu Bradford, for example, sees the EU's expansive set of digital regulations as a kind of "digital constitution" manifesting the continent's commitment to human-centric, rights-preserving, democracy-enhancing, and redistributive policies.[3] We are deeply skeptical that such a diverse body of rules could justify this normative halo. Although rights-based language pervades the EU's rules impacting AI, such as the GDPR and AI Act, its role is managerial, not moral.[4] In the EU, rights are tools of order, not commitments to autonomy or consent.[5] EU AI regulation is steeped in the discourse of rights, but it is not rights-driven or primarily rights-motivated. This distinction is not semantic. In liberal legal systems, rights claims carry legitimating force. To frame EU AI regulation as rights-driven is to cast alternative models as deficient, as if those that prioritize other goals—innovation, federalism, pluralism—fall short of a normative ideal. We offer a different reading: EU regulation is the product of particular historical imperatives, not universal truths.[6] As such it has no claim to universal application.

---

[1] In this Article we focus on AI regulation, a field which overlaps with, but is not merely a subset of, the regulation of "digital markets," and "digital platform regulation." Traditional digital markets regulation primarily governs data privacy and online market dynamics. and Platform regulation adds a layer of governance issues and commonly focuses on issues such as content moderation, algorithmic curation, and automated decision-making processes. AI regulation overlaps with much of this, but also directly engages with how intelligent systems analyze data, generate predictions, and exert influence over human decision-making. Much of this influence is mediated through digital platforms and markets, but not all.

[2] *See,* Anu Bradford, *Whose AI Revolution?*, PROJECT SYNDICATE (Sept. 1, 2023), https://www.project-syndicate.org/onpoint/ai-regulation-us-eu-china-challenges-opportunities-by-anu-bradford-2023-09.

[3] Anu Bradford, *Europe's Digital Constitution*, 64 VA. J. INT'L L.1 (2023). In Part I, *infra*, we show how Bradford's views are emblematic of widespread narrative that the European Union's AI regulatory regime is rooted in a principled commitment to rights.

[4] See, *infra* Part I.

[5] Other scholars have made a similar observation, that EU AI laws are essentially a version of risk-regulation, but without challenging the normative or prescriptive force of rights-based rhetoric. See for example, Margot E. Kaminski, *Regulating the Risks of AI* 103 B.U.L. REV. 1347 (2023); Margot E. Kaminski, *The Developing Law of AI Regulation: A Turn to Risk Regulation*, LAWFARE, (April 21, 2023, 1:23pm), https://www.lawfaremedia.org/article/the-developing-law-of-ai-regulation-a-turn-to-risk-regulation. See also Tobias Mahler, *Between Risk Management and Proportionality: The Risk-Based Approach in the EU's Artificial Intelligence Act Proposal*, 13 EUR. J. RISK REGUL. 120 (2022) (analyzing the EU's AI Act proposal, highlighting its risk-based approach and the balance between risk management and the protection of fundamental rights).

The standard narrative draws a familiar contrast. Europe is said to protect dignity, autonomy, and human oversight. The United States, by contrast, is cast as the domain of market logic and regulatory minimalism. This framing has intuitive appeal. A polity that champions the right to explanation,[7] the right to contest automated decisions,[8] the right to human oversight,[9] the right to be forgotten,[10] and now the right to opt out of AI systems,[11] surely takes rights seriously.[12] But this appeal is largely rhetorical. We do not deny the legal reality of these provisions. We question their role. In practice, they function less as constitutional guarantees than as administrative protocols, invoked to constrain asymmetry and structure discretion. They reflect not a philosophy of law, but a technocratic style of governance oriented toward risk control and institutional preservation.

The European approach to AI regulation mirrors the broader logic of European integration, which has long prioritized balance, containment, and regulatory coordination. From the settlement of Westphalia[13] to the Stability and Growth Pact,[14] the European legal order has been organized around the management of fragmentation and the prevention of systemic disruption. The EU's approach to AI regulation follows this pattern. Its emphasis on rights masks a deeper logic of equilibrium. In contrast, the American model reflects a different trajectory—decentralized authority, constitutional individualism, and an enduring suspicion of bureaucratic overreach. Where Europe builds regulatory coherence through centralized constraint, the United States defaults to pluralism, even at the cost of institutional disarray.

[6] *See, infra* Part I.

[7] *A Right to Explanation*, THE ALAN TURING MACHINE https://www.turing.ac.uk/research/impact-stories/a-right-to-explanation (last visited Feb. 9. 2025).

[8] *See generally* Margot E. Kaminski & Jennifer M. Urban, *The Right to Contest AI*, 121 COLUM. L. REV. 1957 (2021).

[9] *See* Lars Enqvist, *"Human Oversight" in the EU Artificial Intelligence Act: What, When and by Whom?*, 15 LAW, INNOVATION & TECH. 508 (2023), https://doi.org/10.1080/17579961.2023.2245683.

[10] Jeffrey Rosen, *The Right to Be Forgotten*, 64 STAN. L. REV. ONLINE 88 (2012).

[11] *See* Zach Warren, *Legalweek 2024: Current US AI Regulation Means Adopting a Strategic – and Communicative – approach,* THOMSON REUTERS (Feb. 11, 2024), https://www.thomsonreuters.com/en-us/posts/legal/legalweek-2024-ai-regulation/.

[12] For additional EU rights language, see e.g., Proposal for a Regulation of the European Parliament and of the Council Laying Down Harmonised Rules on Artificial Intelligence (Artificial Intelligence Act) and Amending Certain Union Legislative Acts, COM (2021) 206 final, at Art.9 (Apr. 21, 2021) (hereinafter "EU AI Act") (requiring companies to assess potential rights infringements and to mitigate risks). See also Regulation (EU) 2022/2065 of the European Parliament and of the Council of 19 October 2022 on a Single Market for Digital Services and Amending Directive 2000/31/EC (Digital Services Act), 2022 O.J. (L 277), at Art.14 (hereafter, "DSA") (platforms must respect users' rights to freedom of expression, setting out obligations for transparency around content moderation); DSA at Art.23 and 24 (right to transparency and fairness: platforms must be transparent about algorithms, allowing users to understand how content is curated and targeted); DSA Art.17 (Users can contest content moderation decisions, enforcing the right to due process and transparency in online interactions). See also Regulation (EU) 2022/1925 of the European Parliament and of the Council of 14 September 2022 on Contestable and Fair Markets in the Digital Sector (Digital Markets Act), 2022 O.J. (L 265) 1, at Art. 6 (hereinafter "DMA") (data portability and interoperability - users have the right to move data across platforms).

Our aim is not to endorse one model over the other. Both approaches are contingent, and both carry tradeoffs. But much of the perceived legitimacy of the EU's regulatory regime rests on the claim that it is morally superior, because it is rights-based. That claim demands scrutiny. Once we reject the assumption that rights serve as first principles rather than administrative tools, the case for EU normative leadership becomes less persuasive. Its model may be more comprehensive, but its generalizability must be earned through institutional analysis, not assumed through moral branding. The Article proceeds as follows. Part I reconstructs the dominant academic narrative that casts EU AI regulation as a rights-based project. We focus on Anu Bradford's account of the EU's "human-centric" model, but show that her view is broadly shared across emerging AI law scholarship.[15] We then set out our methodological intervention.[16] Drawing from comparative institutional analysis, we argue that regulatory legitimacy must be assessed by how governance is constituted, authorized, and implemented, not by the presence or absence of abstract normative language. We situate the EU framework within its deeper political and legal history, from the logic of Westphalian restraint to the bureaucratic architecture of contemporary European law. This tradition, we argue, privileges institutional balance over moral aspiration.

In Part II, we operationalize this argument by comparing AI governance in five key domains: data privacy, cybersecurity, healthcare, labor and employment, and disinformation. In each, we show how EU regulation prioritizes administrative control and risk mitigation, often invoking rights to justify technocratic oversight. The American approach, by contrast, reflects a structurally different view of state power, grounded in decentralization, market responsiveness, and constitutional permissiveness. The contrast is not one of rights versus markets, but of competing logics of governance.

---

[13] The 1648 Peace of Westphalia ended the Thirty Years' War, which had ravaged Europe and killed some eight million people. The Peace is conventionally seen as the origin of Westphalian sovereignty, the principle that each state has exclusive sovereignty over its territory. See e.g. Derek Croxton, *The Peace of Westphalia of 1648 and the Origins of Sovereignty*, 21(3) INTERNATIONAL HISTORY REVIEW, 569 (1999).

[14] The Stability and Growth Pact is a framework that enforces fiscal discipline among EU Member States. The Pact sets limits on budget deficits and public debt to prevent fiscal mismanagement that could destabilize the broader EU economy. *See generally*, *Stability and Growth Pact*, EUROPEAN COMMISSION, https://economy-finance.ec.europa.eu/economic-and-fiscal-governance/stability-and-growth-pact_en (last visited Jan. 6, 2025)

[15] *See infra* Part I.

[16] We are not alone in questioning the dominant narrative of EU regulatory supremacy. See e.g. Orly Lobel, *The Law of AI for Good*, 75 FLA. L. REV. 1073 (2023) (criticizing various EU AI regulations (and some Biden era initiatives) as regulatory strategies that almost exclusively address the risks of AI while paying short shrift to its benefits.)

We conclude by examining the implications of this institutional framing for global debates over AI governance. The EU model may offer a workable approach for certain jurisdictions, but it should not be treated as a universal paradigm. Governance must be shaped by context—by the political structures, legal traditions, and normative commitments of each society. The real question is not whether AI regulation reflects abstract universal imperatives, but whether it reflects the will and the interests of the people it purports to govern. Even if EU AI regulation makes sense for the EU, we see no reason why the U.S. would want to copy-paste EU regulations.

## I. THE NORMATIVE HALO OF EUROPEAN AI REGULATION

In this Part, we challenge the prevailing academic narrative that the European Union's AI regulatory regime is rooted in a principled commitment to rights. This view, widely held among legal scholars and policy advocates, treats the EU framework as the gold standard for ethical governance, grounded in human dignity, autonomy, and fairness.[17] Our argument cuts against that mainstream position. We suggest instead that the rights language embedded in EU regulation is best understood as a historically contingent response to the risks of political imbalance, not as a universal moral imperative. Part I.A reconstructs the dominant scholarly account, exemplified by Anu Bradford and others, who cast the EU model as a normative export built on the foundation of rights, ethics, and principled design. Part I.B then sets out our methodological intervention. Our comparative institutional approach focuses the details of AI regulation in practice rather than the abstract appeal of its normative form. Finally, Part I.C situates EU regulation within its longer political history. We trace how European legal order, shaped by religious conflicts, imperial rivalries, and the perpetual management of internal fragmentation, has consistently privileged equilibrium over expansion, and containment over assertion. From this perspective, EU AI regulation is not an instrument of universal ethics, but a mechanism of geopolitical risk management. The rights it invokes function less as shields for individual autonomy than as tools for maintaining systemic balance. It is on these grounds that we question the assumption that the European model is normatively superior or globally exportable.

[17] See *infra*, notes 18-37 and accompanying text.

## A. Belief in Rights-Driven AI Regulation

Much of contemporary AI governance scholarship is organized around the language of rights, ethics, and principled design. Anu Bradford, in *Digital Empires* (2023), characterizes the EU approach as human centric, anchored in a framework that safeguards autonomy, democratic engagement, and fairness.[18] We acknowledge the subtlety of Bradford's account: although she sees the export of EU regulation centered on values privacy, dignity, and fairness through a normative lens, she understands that the export is successful primarily because international corporations comply out of economic necessity.[19] As Bradford explains, multinational companies and small nations chose to comply with EU rules because the EU market is too large to ignore, and maintaining separate regulatory frameworks is prohibitively expensive.[20] Despite this, Bradford and others see the successful export of EU rules on subjects like data privacy as a sign of moral authority and intellectual leadership.[21] Bradford notes that "nearly 150 countries have adopted domestic privacy laws, most of them resembling the EU data protection regime" and suggests a similar Brussels Effect will apply to the EU's antitrust rules, regulation of online hate speech and disinformation, and AI.[22] The promise is that "[t]hese policy fields further entrench the global footprint of the European rights-driven regulatory model."[23] In this telling, legal convergence is not simply a byproduct of market power. It is a normative achievement.[24] The normativity of Bradford's project is also evident in her explanation for why the EU lags so far behind the U.S. in digital technology and services. Whatever Europe's structural shortcomings are, "choosing to regulate the tech industry in the name of safeguarding individual rights and societal freedoms is not where the problem lies."[25]

---

[18] *See generally* ANU BRADFORD, DIGITAL EMPIRES: THE GLOBAL BATTLE TO REGULATE TECHNOLOGY (2023) (hereinafter, Bradford, *Digital Empires*); *see also* Anu Bradford, *Europe's Digital Constitution*, 64 VA. J. INT'L L.1 (2023) (arguing that the EU's expansive set of digital regulations can be viewed as Europe's "digital constitution," which engrains Europe's human-centric, rights-preserving, democracy-enhancing, and redistributive vision for the digital economy into binding law).

[19] *See generally* ANU BRADFORD, THE BRUSSELS EFFECT: HOW THE EUROPEAN UNION RULES THE WORLD 1-424 (Oxford Univ. Press, 1st ed. 2020).

[20] Bradford, *Digital Empires, supra* note 18 (restating the "Brussels Effect" thesis and explaining how economic pressure leads to multinational firms and foreign governments adopting EU standards.)

[21] As Julie Cohen summarizes "She [Bradford] does have opinions about why certain values ought to prevail in these contests, and the book's closing section makes those views clear." Julie E. Cohen, *Who's Rulin' Who?,* LAWFARE, Apr. 16, 2024, (https://www.lawfaremedia.org/article/who-s-rulin-who).

[22] Bradford, *Digital Empires*, *supra* note 18 at 325-326.

[23] *Id.*

[24] *Id.*

[25] Anu Bradford, *The False Choice Between Digital Regulation and Innovation*, 119 NW. U. L. REV. 377, 382 (2024).

This view that EU AI regulation can be best explained as "safeguarding individual rights and societal freedoms," as Bradford puts it,[26] or more succinctly, that it is "rights-driven" has gained considerable traction.[27] We focus on Bradford as the preeminent representative of widely held views, not an outlier.[28] For example, scholars such as Paul Schwartz and Daniel Solove follow the EU centric lead, emphasizing that rights discourse remains central to the EU's regulatory framework.[29] Schwartz, in particular, argues that its model of data protection, grounded in fundamental rights like privacy and informational self-determination, has decisively shaped global norms, compelling non-EU jurisdictions to adopt EU style protections.[30] Unlike Bradford's account of unilateral normative diffusion, Schwartz highlights

[26] Bradford, *Digital Empires*, *supra* note 18 at 370.

[27] *See generally* Ryan Calo, *Artificial Intelligence Policy: A Primer and Roadmap.* 51 U.C. DAVIS L. REV. 399 (2017) (discussing the U.S. market-oriented approach to AI regulation, where innovation and economic growth are prioritized, often at the expense of individual rights and protections seen in other jurisdictions); *See also* Mark MacCarthy, *Fairness in Algorithmic Decision-Making*, BROOKINGS (December 6, 2019), https://www.brookings.edu/articles/fairness-in-algorithmic-decision-making/ (exploring differences in fairness approaches, noting that U.S. AI regulation often leans toward market-driven solutions and self-regulation); Joshua A. Kroll, Joanna Huey, Solon Barocas, Edward W. Felten, Joel R. Reidenberg, David G. Robinson & Harlan Yu, *Accountable Algorithms*, 165 U. PA. L. REV. 633 (2017) (discussing that U.S. data privacy regulations are characterized by a fragmented, sectoral approach with decentralized regulatory authority. It indicates that the US approach, focused on sector-specific laws and enforcement by agencies like the FTC, doesn't provide a comprehensive privacy framework, and often prioritizes market interests and commercial flexibility over broad data protection rights).

[28] See also, Graham Greenleaf, *The Influence of European Data Privacy Standards Outside Europe: Implications for Globalization of Convention,* 108, 2 INT'L DATA PRIVACY L. 68 (2012). For additional examples of scholars characterizing EU regulation as rights-driven. See Ronit Justo-Hanani, *The Politics of Artificial Intelligence Regulation and Governance Reform in the European Union*, 55 POLICY SCIENCE 137 (2022), https://link.springer.com/article/10.1007/s11077-022-09452-8 (exploring the EU's integrated policy to tighten control over AI ensures consumer protection and fundamental rights, reflecting a commitment to human-centric regulation); Patricia Gomes Rêgo de Almeida, Carlos Denner dos Santos & Josivania Silva Farias, *Artificial Intelligence Regulation: A Framework for Governance*, 23 ETHICS & INFO. TECH. 505 (2021), https://dl.acm.org/doi/10.1007/s10676-021-09593-z (developing a conceptual framework for AI regulation, emphasizing the importance of embedding ethical considerations and fundamental rights into governance structures); Tambiama Madiega, *EU Guidelines on Ethics in Artificial Intelligence: Context and Implementation,* EUROPEAN PARLIAMENT (2019), https://www.europarl.europa.eu/RegData/etudes/BRIE/2019/640163/EPRS_BRI(2019)640163_EN.pdf (providing an overview of EU's guidelines on AI ethics, showing a commitment to human-centric and rights-based approaches in AI development and deployment); Alessandro Mantelero, *AI and Big Data: A Blueprint for a Human Rights, Social and Ethical Impact Assessment,* 4 J.CYBERSECURITY PRIVACY 43 (2022), https://www.mdpi.com/2571-8800/4/4/43 (discussing the EU's AI Act proposal, focusing on its implication for consumer protection and fundamental rights, showing a rights-driven regulatory approach).

[29] *See* Daniel J. Solove and Paul Schwartz, *Reconciling Personal Information in the United States and European Union*, 102 CALIF. L. REV. 877 (2014) (arguing that in the US, privacy law focuses on redressing consumer harm and balancing privacy with efficient commercial transactions while in the EU, privacy is hailed as a fundamental right).

[30] *See generally* Paul M. Schwartz, *Global Data Privacy: The EU Way*, 94 N.Y.U L. REV 771 (2019) (emphasizing that the discourse of rights remains central to the EU's regulatory framework, arguing that its data protection model, grounded in fundamental rights like privacy and informational self-determination, has decisively shaped global

the EU's reliance on bilateral agreements, adequacy decisions, and institutional pathways to extend its influence.[31] Still, the focus remains on rights as the engine of legitimacy, as if the vocabulary alone guarantees the normative content.[32]

Woodrow Hartzog and Neil Richards similarly defend the EU's rights oriented framework, contrasting it with the United States' more permissive, market driven approach.[33] In their account, American privacy law is designed to redress consumer harm and balance privacy against commercial efficiency.[34] The European model, by contrast, treats privacy as a constitutional commitment.[35] Under the GDPR, dignity and autonomy are not market considerations.[36] They are the foundation. Accordingly, they argue, the United States should learn from Europe and adopt stronger rights based protections to restore human dignity in the digital age.[37]

---

norms, compelling non-EU jurisdictions to adopt EU-style protections. However, unlike Bradford's unilateralism, Schwartz highlights the EU's reliance on bilateral).

31 Paul M. Schwartz, *The EU-U.S. Privacy Collision: A Turn to Institutions and Procedures*, 126 HARV. L. REV. 1966, 1994-1997 (2013) (examining the differences between US and EU data protection frameworks, criticizing the U.S. approach for its limited scope and emphasizes the EU's rights-based model as a more comprehensive standard. He suggests that the US could enhance its privacy framework by integrating elements of the EU's approach)

32 For the other literature, *see generally* Daniel J. Solove, *Understanding Privacy* (Harvard Univ. Press 2010) (exploring the foundational principles of privacy law and criticizing the fragmented nature of US privacy regulations. He demonstrates the influence of the EU, through GDPR, on global privacy standards, suggesting that the GDPR's comprehensive approach to privacy could serve as a blueprint for the US to establish a more uniform and robust privacy framework. It also implies that the US would benefit from a shift toward a right-based approach to privacy.); *see also* Paul M. Schwartz & Karl-Nikolaus Peifer, *Transatlantic Data Privacy*, 106 GEO. L.J. 115 (2017) (discussing the philosophical and procedural divergences between U.S. and EU privacy laws, emphasizing how these differing approaches construct unique "legal identities" around data privacy, suggesting that the US could benefit from borrowing aspects of the EU's regulatory framework, particularly by adopting collaborative "harmonization networks" and mutual recognition mechanisms to bridge legal and cultural differences in data privacy.); David Cole & Federico Fabbrini, *Bridging the Transatlantic Divide? The United States, the European Union, and the Protection of Privacy Across Borders, Courts Working Paper Series No. 33* (Nov. 20, 2015) (comparing US and EU privacy protections, noting that the EU's rights-based model offers a robust framework that the US could consider, particularly in light of shared concerns over government surveillance).

33 Woodrow Hartzog & Neil M. Richards*, Privacy's Constitutional Moment and the Limits of Data Protection,* 61 B.C. L. REV. 1687, 1690-1696 (2020) (discussing the differences between the U.S. consumer protection framework and the EU's rights-focused approach under the GDPR, noting that the U.S. framework is more permissive and focused on consumer vulnerabilities rather than individual rights).

34 *Id.*

35 *Id.*

36 *Id.*

37 *Id.*

Beyond rights protection, ethics is another cornerstone of the prevailing discourse. The High Level Expert Group on AI, appointed by the European Commission, foregrounded ethics in its 2019 *Ethics Guidelines for Trustworthy AI*.[38] It identified seven key requirements: human agency and oversight, technical robustness and safety, privacy and data governance, transparency, diversity and non-discrimination, societal and environmental wellbeing, and accountability.[39] These principles have since been incorporated into major EU legislative proposals, including national AI strategies in Germany,[40] France,[41] and the Netherlands.[42]

This ethical leadership is widely celebrated in academic circles. Scholars such as Luciano Floridi and Virginia Dignum, writing in *Minds and Machines* and *AI and Society*, praise the EU's ethical framework as the normative core of AI governance.[43] The goal, they argue, is not merely to prevent harm but to promote human flourishing.[44] Ethics, in this view, is equivalent to the source of legitimacy.[45] Legal structures are justified not because they constrain power but because they instantiate moral substance.

The third prong is principled design. Referring to the effort to embed normative commitments directly into the architecture of algorithmic systems, this approach builds on Batya Friedman's theory of value sensitive design, first developed in the 1990s.[46] The core claim of this school is that fairness, accountability, and transparency are not external legal constraints to be applied

---

[38] *See Ethical Guidelines for Trustworthy AI*, EUROPEAN COMMISSION (April 8, 2019), https://digital-strategy.ec.europa.eu/en/library/ethics-guidelines-trustworthy-ai.

[39] *Id.*

[40] *See* Alena Buyx, *AI - its Risks and Possible Side Effects*, DEUTSCHLAND.DE (April 4, 2023), https://www.deutschland.de/en/topic/culture/german-ethics-council-risks-of-artificial-intelligence.

[41] *See Global/France: AI Action Summit Must Meaningfully Center Binding and Enforceable Regulation to Curb AI-Driven Harms*, AMNESTY INTERNATIONAL (February 9, 2025), https://www.amnesty.org/en/latest/news/2025/02/global-france-ai-action-summit-must-meaningfully-center-binding-and-enforceable-regulation-to-curb-ai-driven-harms/.

[42] *See e.g.,* Eileen Guo, Gabriel Geiger, and Justin-Casimir Braun, *Inside Amsterdam's High-Stakes Experiment to Create Fair Welfare AI*, MIT TECHNOLOGY REVIEW (June 11, 2025), https://www.technologyreview.com/2025/06/11/1118233/amsterdam-fair-welfare-ai-discriminatory-algorithms-failure/.

[43] *See generally*, Luciano Floridi et al., *AI4People - An Ethical Framework for a Good AI Society: Opportunities, Risks, Principles and Recommendations*, 28 MINDS & MACHINESK 689 (2018).

[44] *Id.*

[45] *Id.* (only implicitly, however, as the text strongly suggests that ethical acceptability is a precondition for public trust, adoption, and governance legitimacy)

[46] *See generally,* Batya Friedman, *Value-Sensitive Design*, 3 ACM INTERACTIONS 16 (Dec. 1996)

after the fact.[47] They are design criteria to be embedded from the outset.[48] Within the EU, this design philosophy has been institutionalized through regulatory requirements.[49] Article 9 of the EU AI Act, for instance, requires providers of high risk systems to establish and maintain quality management systems that track ethical compliance throughout the product lifecycle.[50]

This logic has begun to spread beyond Europe, even briefly in the United States. Most notably, the 2023 White House *Blueprint for an AI Bill of Rights* laid out five principles: safe and effective systems, algorithmic discrimination protections, data privacy, notice and explanation, and human alternatives.[51] These mirror the EU's rights ethics design triad. In 2024, the Office of Management and Budget issued Memorandum M 24 10, requiring federal agencies to inventory, assess, and mitigate risks from AI systems in high impact domains.[52] Under the Biden Administration there were signs that the U.S. was moving closer to Europe on AI regulation, but that course was quickly reversed by the incoming Trump Administration. The Biden administration's approach to AI governance was characterized by an emphasis on comprehensive safety measures, regulatory frameworks, and equitable deployment considerations.[53] This philosophy was most prominently embodied in Executive Order 14110, which established extensive oversight mechanisms designed to ensure the "safe, secure, and trustworthy development and use of artificial intelligence."[54] However, the Trump administration wasted no time in announcing a decisive policy reversal through Executive Order 14179, "Removing Barriers to American Leadership in Artificial Intelligence," announced on day one of the new administration.[55] This executive order signaled a fundamental departure from safety-centric regulation toward innovation-driven competition.[56] Whereas the Biden Administration focused on issues relating to AI safety, regulatory oversight, civil rights and equity, the Trump administration is focused on promoting American technology dominance by removing barriers to innovation, building out America's AI infrastructure, and promoting similar policies overseas.[57]

[47] *Id. See also Value Sensitive Design*, AI ETHICS LAB https://aiethicslab.rutgers.edu/e-floating-buttons/value-sensitive-design/ (last visited July 30, 2025).

[48] *Id.*

[49] *See* AJ Richter, *Ethical AI: How Data Officers Craft Policies for Fairness, Accountability, and Transparency*, TECHGDPR (October 16, 2024), https://techgdpr.com/blog/ethical-ai-how-data-officers-craft-policies-for-fairness-accountability-and-transparency/.

[50] *See Article 9: Risk Management System*, EU AI ACT (August 2, 2026), https://artificialintelligenceact.eu/article/9/.

[51] *See Blueprint for an AI Bill of Rights*, THE WHITE HOUSE, https://bidenwhitehouse.archives.gov/ostp/ai-bill-of-rights/ (last visited July 30, 2025).

[52] *See Memorandum for the Heads of Executive Departments and Agencies*, EXECUTIVE OFFICE OF THE PRESIDENT (March 28, 2024), https://www.whitehouse.gov/wp-content/uploads/2024/03/M-24-10-Advancing-Governance-Innovation-and-Risk-Management-for-Agency-Use-of-Artificial-Intelligence.pdf.

[53] See generally EXECUTIVE ORDER 14110, SAFE, SECURE, AND TRUSTWORTHY DEVELOPMENT AND USE OF ARTIFICIAL INTELLIGENCE, 88 Fed. Reg. 75,191 (Oct. 30, 2023).

[54] *Id.*

[55] EXECUTIVE ORDER 14179, REMOVING BARRIERS TO AMERICAN LEADERSHIP IN ARTIFICIAL INTELLIGENCE (Jan. 20, 2025), available at https://www.whitehouse.gov/presidential-actions/2025/01/removing-barriers-to-american-leadership-in-artificial-intelligence/.

Across all three prongs—rights, ethics, and principled design—the underlying logic remains the same. Because rights are invoked, ethics articulated, and technical systems built to reflect these values, the regulatory system itself is presumed to be not only effective but just. The implication is that legitimacy follows naturally from design. And by extension, systems that do not embed such values must necessarily fail, whether institutionally, morally, or legally.[58] What results is a theory of governance in which legitimacy is presumed to arise from normative form, not from political origin.

But this understanding of regulatory legitimacy obscures more than it reveals. Just because rights and values are invoked does not mean the substance they claim to represent is actually present. History offers no shortage of examples in which the language of rights was used to accomplish the very things those rights were supposed to prevent.[59] Francisco de Vitoria, in his sixteenth century *Relectio de Indis*, invoked the *ius peregrinandi*, the right to travel, and the *ius praedicandi*, the right to preach Christianity, as legal grounds for Spanish presence in the Americas.[60] Should Indigenous peoples refuse these rights, he argued, they could be lawfully compelled by force. As he wrote: If the Indians were to refuse to allow the Spaniards to preach

---

[56] *Id.* The administration's vision for AI has been further articulated in the 2025 "AI Action Plan," released under the title "Winning the AI Race: America's AI Action Plan." WHITE HOUSE, WINNING THE AI RACE: AMERICA'S AI ACTION PLAN (2025), available at https://www.ai.gov/action-plan. (outlining over ninety federal policy actions organized around three core pillars: accelerating innovation, building American AI infrastructure, and leading in international diplomacy and security. White House Unveils America's AI Action Plan, WHITE HOUSE (2025), https://www.whitehouse.gov/briefing-room/statements-releases/2025/01/ai-action-plan/.

[57] *Id.*

[58] *See* Ben Dubow, *Part One: Watch Out Russia and China: AI is a Threat*, CEPA (August 1, 2023), https://cepa.org/article/part-one-watch-out-russia-and-china-ai-is-a-threat/ (Arguing that China's rules intended to keep AI bound by "core socialist values" will limit the country's AI technological advances).

[59] *See generally* SAMUEL MOYN, THE LAST UTOPIA (Harvard Univ. Press 2011) (ebook) (showing that human rights become prominent after the collapse of alternative political utopias, criticizing the idea that rights are inherently liberatory, and how they often serve as a minimal substitute for political transformation).

[60] *See generally* ANTHONY PAGDEN & JEREMY LAWRENCE, FRANCISCO DE VITORIA: POLITICAL WRITINGS (Cambridge Univ. Press 1991) (saying that the Spaniards may lawfully travel and dwell among the barbarians... the right of travel and sojourn cannot be denied to innocent people... [the Europeans] also have the right to announce and preach the Gospel in the lands of the barbarians)

the Gospel, they would be committing a mortal sin... and might be compelled by war.[61] Vitoria's logic, cloaked in the language of justice, civility, and natural law, set the intellectual groundwork for centuries of colonial jurisprudence, in which rights were not shields for the weak but instruments of domination for the strong.[62]

And in today's regulatory imagination, who is to say the same dynamic is not quietly at work? For example, the newly enacted UK Online Safety Act aims to protect minors by requiring age verification for certain internet activities. However, this child safety measure may erode adult privacy by requiring additional data collection and limiting options for those who wish to browse and post anonymously.[63] Users who seek to protect their privacy by the use of VPNs may also be in a bind as those providers have been known to sell user data to data brokers.[64] A similar irony can be seen in with respect to the GDPR. Whatever the salutary effects of the GDPR might be, a substantial body of evidence shows that the regulation has reduced overall website traffic by imposing costs that are disproportionately felt by smaller platforms.[65] Likewise in relation to app development, Rebecca Janßen and coauthors concluded that "[w]hatever the benefits of GDPR's privacy protection, it appears to have been accompanied by substantial costs to consumers, from a diminished choice set, and to producers from depressed revenue and increased costs."[66] In addition, it has been observed that by restricting the use and collection of sensitive personal data (like race, ethnicity or religion), the GDPR may frustrate efforts to auditing AI systems for bias.[67] Normatively infused language should not be used to justify avoiding hard questions about practical effects. Given this, what is needed is not more invocations of rights, values, ethics, or any other morality flavored terms, but a structural account of governance and a pragmatic assessment of how regulations actually function in practice.

---

[61] *Id* at ch. 6 (on the American Indians).

[62] *See* Antony Anghie, *Francisco de Vitoria and the Colonial Origins of International Law*, 5 SOC. & LEGAL STUD. 321 (1996).

[63] Matt Burgess & Lily Hay Newman, *The Age-Checked Internet Has Arrived*, WIRED (July 25, 2025), https://www.wired.com/story/the-age-checked-internet-has-arrived/.

[64] *Id.*

[65] Julia Schmitt, Klaus M. Miller & Bernd Skiera, *The Impact of Privacy Laws on Online User Behavior*, ARXIV (Oct. 19, 2021) (revised version), arXiv:2101.11366 [econ.GN].

[66] Rebecca Janßen, Reinhold Kesler, Michael E. Kummer, and Joel Waldfogel, *GDPR and the Lost Generation of Innovative Apps*, Working Paper No. w30028 (Nat'l Bureau of Econ. Research, 2022),(https://www.nber.org/system/files/working_papers/w30028/w30028.pdf).

[67] Marvin van Bekkum & Frederik Zuiderveen Borgesius, *Using Sensitive Data to Prevent Discrimination by Artificial Intelligence: Does the GDPR Need a New Exception?,* ARXIV, (Nov. 28, 2022 revision) (2022), arXiv:2206.03262 [cs.CY].

## B. The Method: Comparative Institutional Analysis

To answer these questions, this Article adopts a comparative institutional method. The aim is not to adjudicate between competing rights claims, nor to declare whether the European or American model is more democratic, efficient, or just. It does not attempt to reconstruct rights on new philosophical grounds. Instead, it seeks to understand how each system governs in practice, through institutional design, administrative structure, and regulatory implementation.

As a result, we examine how regulations are produced, how historical forces shape legal frameworks, and why certain regulatory models gain global influence. By examining these frameworks' institutional and historical contexts, this Article challenges assumptions about European regulatory superiority and universal applicability.

## C. The Reality of Historically Contingent Regulation

The European regulatory setup is historically contingent, as its order has long been premised on maintaining a delicate balance of power between states, constraining overreach and mitigating systemic risks.[68] The Peace of Westphalia in 1648—which ended the Thirty Years' War and marked the birth of modern state sovereignty—rejected domination by empires and religious authorities.[69] This Westphalian system established a core principle: no state could impose its will on another; each would govern itself, free from external interference.[70] It set the model for European coexistence—a calculated balance against the risks of unchecked ambition and domination.[71]

[68] *See generally* ANDREW MORAVCSIK, THE CHOICE FOR EUROPE: SOCIAL PURPOSE AND STATE POWER FROM MESSINA TO MAASTRICHT (Cornell Univ. Press 1998) (Arguing that the motivation behind European integration is a calculated balance of power among states); Tanja Börzel & Thomas Risse, *When Europe Hits Home: Europeanization and Domestic Change*, 4 EUR. INTEGRATION ONLINE PAPERS ,https://eiop.or.at/eiop/pdf/2000-015.pdf, (arguing that Europeanization is fundamentally about balancing local interests with the benefits of collective risk management); Alec Stone Sweet & Thomas L. Brunell, *Constructing a Supranational Constitution: Dispute Resolution and Governance in the European Community*, 92 AM. POL. SCI. REV. 63 (1998) (analyzing the role of European legal institutions in balancing state sovereignty with supranational governance, particularly through the European Court of Justice, showing how legal frameworks are designed to mitigate cross-border risks while respecting the integrity of member states); Kalypso Nicolaïdis, *We, the Peoples of Europe…*, 83 FOREIGN AFFAIR. 97 (2004) (arguing for a "democratic European order" that balances integration with local autonomy and emphasizes risk management in areas like security and economic stability).

[69] *See generally* ANDREAS OSIANDER, THE STATES SYSTEM OF EUROPE, 1640–1990: PEACEMAKING AND THE CONDITIONS OF INTERNATIONAL STABILITY (Oxford Univ. Press 1994) (analyzing the Peace of Westphalia and its significant influence on the development of the European state system, arguing that the treaties established principles of state sovereignty and mutual recognition, effectively ending imperial and religious dominance in Europe. This This marked a transition toward a system of coexisting, sovereign entities, setting a precedent that influenced international relations and diplomacy in Europe and rest of the world)

[70] *See e.g.,* KALEVI J. HOLSTI, PEACE AND WAR: ARMED CONFLICTS AND INTERNATIONAL ORDER, 1648-1989 (Cambridge Univ. Press 1991) (examining the evolution of international order, beginning with the Peace of Westphalia, arguing that Westphalia established foundational principles of state sovereignty and non-intervention.)

This balancing act wasn't merely philosophical; it was an existential strategy.[72] The unchecked expansion of one state could destabilize the broader system and threaten everyone.[73] Russia, for instance, often challenged this balance with its relentless, messianic drive for expansion.[74] When it pushed into Central Asia in the 19th century, it ignited the "Great Game" with Britain, drawing their European allies—France and the Ottoman Empire for Britain, Prussia and Austria-Hungary for Russia—into a broader conflict.[75] The interconnected alliances meant that tensions in distant regions could easily escalate into major European struggles; if one state disrupted the order, the entire system faltered.[76] To maintain peace in Europe, therefore, each state must temper the ambitions of others—whether by granting autonomy, enforcing mutual respect, or building strategic alliances.[77]

This orientation toward equilibrium remains embedded in the institutional DNA of the EU. Maastricht Treaty enshrined the principle of subsidiarity, mandating that the EU act only when objectives cannot be effectively achieved by member states.[78] The Stability and Growth Pact imposes fiscal discipline to prevent domestic mismanagement from endangering the monetary

---

71 *See* Leo Gross, *The Peace of Westphalia, 1648–1948*, 42 AM. J. INT'L L. 20 (1948) (arguing that the Peace of Westphalia is designed to facilitate peaceful coexistence among independent states and prevent domination by any one entity)

72 *See generally* KENNETH N. WALTZ, THEORY OF INTERNATIONAL POLITICS (Waveland Press, Inc. 2010) (arguing that states pursue balance of power as a survival mechanism in an anarchic international system, explaining that when one state becomes too powerful, others align to counterbalance it, as unchecked expansion threatens the stability and security of all states).

73 *Id.*

74 *See generally* PAUL W. SCHROEDER*,* THE TRANSFORMATION OF EUROPEAN POLITICS, 1763-1848 (Oxford Univ. Press 1994) (highlighting how Russia's drive for territorial and ideological expansion was perceived as a destabilizing force).

75 HENRY KISSINGER, CHAPTER 8 in DIPLOMACY (1994).

76 For example, in the mid 19th century, Russia sought to expand its influence over the Black Sea region. It tried to seize Ottoman territories, particularly Crimea, and gain control over the Bosporus and Dardanelles straits. These actions directly led to the Crimean War (1853-1856), with Britain, France, and Sardinia intervening to prevent Russian dominance.

77 *See generally* RICHARD LITTLE, THE BALANCE OF POWER IN INTERNATIONAL RELATIONS: METAPHORS, MYTHS AND MODELS (2007).

78 *See The Principle of Subsidiarity*, FACT SHEETS ON THE EUROPEAN UNION, https://www.europarl.europa.eu/factsheets/en/sheet/7/the-principle-of-subsidiarity (last visited Feb 10, 2025) union.[79] The Free Movement Directives protect economic integration while restricting measures that might fragment the common market.[80] Even the European Arrest Warrant, which facilitates judicial cooperation, operates within a structure that preserves national sovereignty.[81] These are not expressions of ideological harmonization. They are instruments of political restraint.

The same logic underlies the European Union's use of rights language. In Europe, rights emerged not as universal abstractions but as political tools forged in the crucible of sectarian violence.[82] During the French Wars of Religion, Étienne de La Boétie cast the recovery of natural rights as a refusal to submit to tyranny.[83] Pamphleteers like Théodore de Bèze defended armed resistance by invoking natural rights, not as metaphysical principles, but as bulwarks against annihilation.[84] In this context, rights were not symbols of Enlightenment optimism. They were tactical mechanisms for restoring order in divided societies.[85]

Given this genealogy, it is a mistake to read the EU's invocation of "rights" in the AI context as a principled commitment to individual autonomy. More often, it reflects a deeper concern: the disruptive potential of technological asymmetry. AI is not a neutral tool. It consolidates power, scales decision-making, and introduces new geopolitical vulnerabilities.[86] Should any one state gain strategic dominance in this domain,[87] it could undermine the fragile equilibrium upon which the European project depends.[88] Framing AI regulation in terms of rights serves not only to signal ethical leadership, but to preempt the systemic risks that such asymmetries would entail.[89]

---

[79] *See generally*, *Stability and Growth Pact*, EUROPEAN COMMISSION, https://economy-finance.ec.europa.eu/economic-and-fiscal-governance/stability-and-growth-pact_en (last visited Jan. 6, 2025).

[80] *See Free Movement of Capital,* FACT SHEETS ON THE EUROPEAN UNION, https://www.europarl.europa.eu/factsheets/en/sheet/39/free-movement-of-capital (last visited Feb 10, 2025)

[81] *See European Arrest Warrant*, EUROPEAN COMMISSION, https://ec.europa.eu/info/law/cross-border-cases/judicial-cooperation/types-judicial-cooperation/european-arrest-warrant_en (last visited Jan. 6, 2025).

[82] *See e.g.,* DAN EDELSTEIN, ON THE SPIRIT OF RIGHTS 28-31, 34-39 (Univ. of Chi. Press 2018) (describing how people like Théodore de Bèze used natural rights as a barrier against sovereign overreach, and terms like "divine and human rights" specifically in opposition to political subjugation. It also discussed how pamphleteers such as John Lilburne and Richard Overton combined natural rights discourse with constitutional claims to argue against oppression, arguing that these rights were foundational and should restrict government overreach)

[83] *Id.* at 27.

[84] *Id.* at 29.

[85] *See also* Corina Lacatus, *Balancing Legalism and Pragmatism: A Qualitative Content Analysis of Human Rights Language in Peace Agreements*, 16 J. HUM. RTS. PRAC. 325 (2024), https://doi.org/10.1093/jhuman/huad038. *See also Artificial Intelligence (AI) and Human Rights: Using AI as a Weapon of Repression and Its Impact on Human Rights*, THINK TANK EUROPEAN PARLIAMENT (May, 2024), https://www.europarl.europa.eu/thinktank/en/document/EXPO_IDA(2024)754450. (emphasizing that AI misuse can lead to systemic repression, manipulation, and suppression of dissent. This destabilizes international norms and values, creating global tensions. Therefore, it's important to integrate human rights principles into AI governance to prevent the spread of "algorithmic authoritarianism.").

[86] James Pethokoukis, *An Encouraging Study on the Transformative Potential of AI*, AM. ENTER. INST. (Mar. 18, 2024), https://www.aei.org/economics/an-encouraging-study-on-the-transformative-potential-of-ai/.

Hence, European political consciousness remains oriented toward interest-balancing and risk mitigation,[90] precisely because EU identity is predicated on fractured sovereignty.[91] The legitimacy of the EU depends on its capacity to preserve equilibrium, not to project universal rule.[92] Placing AI regulation within this historical and institutional context helps explain why

---

87 See e.g., Barry Pavel et al., AI and Geopolitics: How Might AI Affect the Rise and Fall of Nations, RAND (Nov 3, 2023), https://www.rand.org/pubs/perspectives/PEA3034-1.html.

88 Warren Chin, *Technology, War and the State: Past, Present and Future*, 95 INT'L AFF. 765 (2019), https://doi.org/10.1093/ia/iiz106 (examining the relationship between technological advancements and warfare, showing how innovation often leads to countermeasures and escalate conflicts). A quick look at European history makes this clear. In the 19th century, Britain's navy wasn't just the largest; it was the most technologically advanced. It shifted from traditional sail-powered vessels to steam propulsion. It also launched HMS Warrior in 1860, the Royal Navy's first iron-hulled warship. Its weaponry was significantly improved, including the development of more powerful and accurate guns, as well as pioneering the use of iron in shipbuilding, allowing for the construction of larger and more robust vessels. This dominance allowed Britain to control key territories across Africa, Asia, and the Caribbean. *See generally* Steven J. Holcomb, *A Century of British Dominance of the Mediterranean: Lessons for the U.S. Navy in the South China Sea*, U.S. NAVAL INSTITUTE (June 2021), https://www.usni.org/magazines/naval-history-magazine/2021/june/century-british-dominance-mediterranean-lessons-us-navy. *See also,* N.A.M. RODGER, SEA-POWER AND EMPIRE, 1688–1793*,* IN THE OXFORD HISTORY OF THE BRITISH EMPIRE: THE EIGHTEENTH CENTURY 169–183 (P.J. Marshall ed., 1998), https://doi.org/10.1093/acprof:oso/9780198205630.003.0008 (discussing how British sea power was integral to the empire's expansion during the 18th century).

89 Jared Cohen, *The Next AI Debate Is About Geopolitics*, FOREIGN POLICY (Oct. 28, 2024), https://foreignpolicy.com/2024/10/28/ai-geopolitics-data-center-buildout-infrastructure/.

90 *See generally,* ULRICH BECK, RISK SOCIETY: TOWARDS A NEW MODERNITY (Mark Ritter trans., SAGE Publications 1992) (arguing that modern societies, particularly those in Europe, are fundamentally shaped by a need to manage complex, "manufactured" risks—risks arising from technological advancements, environmental crises, and industrial hazards. They have evolved into "risk societies," where governments and institutions proactively seek to anticipate, manage, and mitigate risks. Additionally, the "individualization" of risk has prompted European governments to take on a role that balances personal freedoms with the need for communal security).

91 *See generally*, ANTHONY PAGDEN, ET. AL*.,* THE IDEA OF EUROPE: FROM ANTIQUITY TO THE EUROPEAN UNION (2002) (tracing the continent's evolution as a patchwork of independent political entities, from ancient Greece and Rome to the modern European Union. He argues that, historically, Europe never coalesced into a singular, unified political entity; rather, it developed as a collection of diverse sovereignties, each with distinct cultures, languages, and governing structures).

92 *See generally* JÜRGEN HABERMAS, THE POSTNATIONAL CONSTELLATION: POLITICAL ESSAYS (Max Pensky trans., 2001) (arguing that Europe's structure, particularly within the European Union, represents a move away from centralized national sovereignty toward a system of shared governance that respects the autonomy of individual states while fostering a collective European identity); ERNST B. HAAS, THE UNITING OF EUROPE: POLITICAL, SOCIAL, AND ECONOMIC FORCES, 1950–1957 (rev. ed. 2004) (arguing that European integration operates through a process of "functional spillover," where cooperation in specific areas creates incentives for references to rights, even when sincerely held, may function more as instruments of stability than as expressions of moral universality. It also sharpens our skepticism that the EU's regulatory model—however well-articulated—should be treated as normatively superior or universally exportable. In Part II, we put that skepticism into action.

## II. THE EU/US DIVIDE ON AI REGULATION IN FIVE CASE STUDIES

Until now, we have reconstructed the prevailing academic view that casts the European Union's AI regulatory regime as grounded in a principled commitment to fundamental rights, and we have raised reasons for skepticism. In this Part, we carry that skepticism forward through comparative institutional analysis. We examine five domains: data privacy, cybersecurity, healthcare, labor and employment, and disinformation. In each, the EU and the United States adopt markedly different regulatory approaches. These differences are not merely technical. They reflect distinct political logics. The EU model, shaped by a long tradition of risk containment and power-balancing, uses the language of rights as a tool of administrative order rather than as an affirmation of moral autonomy. The U.S. model, by contrast, prioritizes decentralized governance, market flexibility, and constitutional individualism, even at the cost of systemic incoherence. Across all five domains, what emerges is not a contest between values and markets but between two institutional strategies for managing technological disruption. These case studies do not attempt to vindicate one model or reconstruct rights on more persuasive grounds. They aim instead to explain why the EU model, however comprehensive or well-designed, should not be treated as normatively superior by default. It is historically contingent, shaped by specific political conditions, and should not be universalized. Other countries should design their own regulatory frameworks in light of their institutional traditions, political structures, and constitutional commitments.

### A. Data Privacy and Protection

The popular view that EU AI regulation is principally grounded in a rights-based normative vision owes a great deal to the GDPR and European Privacy law more generally. Privacy law is the area in which we anticipate the most resistance to our contention the European approach to AI regulation is driven by the felt necessity of constraining technological disruption and

---

broader collaboration, eventually encouraging integration across a wider range of policies and economic sectors. This process is incremental and relies on fostering interdependence between sovereign states rather than imposing a centralized authority.) *See also* Jan-Werner Müller, *A General Theory of Constitutional Patriotism*, 6 INT'L J. CONST. L. 72 (2008) (discussing that in Europe where nations retain distinct identities, languages, and political systems, they have a form of constitutional patriotism that fosters a form of attachment rooted in shared principles and democratic values).

imposing centralized risk management, and not a recognition of fundamental universal human rights. As such we address it first.

### *1. A Comparison of European Union and United States Data Privacy Law*

It is all too easy to draw and contrast between the EU and US approaches to privacy law in conventional terms. In the EU privacy rights are strongly protected through a one-size fits all continent-wide regime with real enforcement mechanisms. EU citizens are given extensive rights in relation to the collection and processing of their personal information by a broad set of data processors.[93] These rights are enforced within an administrative structure where National Data Protection Authorities (NDPAs) play a central role, alongside the European Data Protection Authorities (EDPB) and national courts.[94] As independent public entities established in each EU member state, NDPAs oversee GDPR compliance, adjudicate individual complaints, and enforce sanctions against violators.[95] Their authority extends to conducting investigations, issuing warnings, mandating cessation of unlawful data processing, and imposing significant financial penalties on noncompliant organizations.[96]

A closer examination of the GDPR's provisions is essential to fully appreciate its legal and practical significance. Under the GDPR, organizations processing the data of EU residents, regardless of their geographic location, must comply with its requirements or face severe penalties.[97] These penalties can reach up to €20 million or 4% of global annual revenue, whichever is higher. [98] The GDPR's enforcement is not symbolic: British Airways was fined £20 million for a data breach; and Google was fined €50 million by France's CNIL for inadequate transparency in targeted advertising.[99] Compliance mechanisms include the appointment of Data Protection Officers (DPOs), [100] who oversee data protection strategies, conduct audits, and serve as liaisons with regulatory authorities and data subjects.[101] Organizations must also conduct Data Protection Impact Assessments (DPIAs) to evaluate risks to individual rights, document mitigation measures, and ensure data processing adheres to principles of transparency, necessity, and proportionality.[102] Furthermore, technical safeguards like encryption, pseudonymization, and secure storage are mandatory to protect personal data and demonstrate compliance.

[93] For a general introduction of GDPR, *See Considering Data Protection and Privacy (GDPR)*, EUROPEAN INSTITUTE OF MANAGEMENT AND FINANCE, https://eimf.eu/considering-data-protection-and-privacy-gdpr/ (last visited Jan. 6, 2025).

[94] *See Data Protection Authority & You*, DATA PROTECTION GUIDE FOR SMALL BUSINESS https://www.edpb.europa.eu/sme-data-protection-guide/data-protection-authority-and-you_en?utm_source=chatgpt.com (last visited Feb.9, 2025).

[95] *Id.*

[96] *Id.*

[97] *See e.g., Data Protection Under GDPR*, YOUR EUROPE, https://europa.eu/youreurope/business/dealing-with-customers/data-protection/data-protection-gdpr/index_en.htm (last visited Nov. 15, 2024) (saying that organizations must clearly inform individuals about who is processing their personal data and why; organizations are also responsible for complying with all data protection principles and must demonstrate this compliance. )

[98] *Id* (companies based in EU or companies established outside the EU but processes personal data in relation to the offering of goods or services to individuals in the EU must comply with GDPR).

[99] *See ICO Fines British Airways £20m for Data Breach Affecting More Than 400,000 Customers*, GDPR REGISTER https://www.gdprregister.eu/news/british-airways-fine/ (last visited Nov 15, 2024); *CNIL's Restricted Committee*

Individuals, under GDPR, are granted the right to access, correct, and delete their personal data, as well as the right to data portability.[103] They can object to specific data processing activities, such as direct marketing, and restrict processing under certain conditions, as when data accuracy is in dispute.[104] Explicit, informed consent must be provided for their data to be processed and used.[105] As to regulatory authorities, the GDPR ensures strict compliance across the EU.[106] As mentioned, each Member State must establish an independent Data Protection Authority (DPA) responsible for overseeing GDPR adherence, investigating complaints, imposing penalties for violations, and supervising cross-border data transfers.[107] To uphold EU data protection standards internationally, regulatory authorities also rely on Standard

---

*Imposes a Financial Penalty of 50 Million Euros Against Google*, EUROPEAN DATA PROTECTION BOARD (Jan. 21, 2019), https://www.edpb.europa.eu/news/national-news/2019/cnils-restricted-committee-imposes-financial-penalty-50-million-euros_en.

100 *See Understanding the Role of Data Protection Officers (DPOs)*, EGNYTE, https://www.egnyte.com/guides/governance/dpo. (last visited Nov. 15, 2024)

101 *Data Protection Officer (DPO)*, EUROPEAN DATA PROTECTION SUPERVISOR, https://www.edps.europa.eu/data-protection/data-protection/reference-library/data-protection-officer-dpo_en (last visited Nov. 15, 2024).

102 *See Data Protection Impact Assessments (DPIAs)*, DATA PROTECTION COMMISSION, https://www.dataprotection.ie/en/organisations/know-your-obligations/data-protection-impact-assessments (last visited Nov. 15, 2024).

103 *See Rights of the Individual*, EUROPEAN DATA PROTECTION SUPERVISOR, https://www.edps.europa.eu/data-protection/our-work/subjects/rights-individual_en (last visited Nov.15, 2024).

104 *Id.*

105 *See What are the Consent Requirement*, GDPR.EU, https://gdpr.eu/gdpr-consent-requirements/ (last visited Nov. 15, 2024).

106 *See e.g., European Commission Press Release: Stronger Rules on Data Protection in the EU*, EUROPEAN COMMISSION, https://ec.europa.eu/commission/presscorner/detail/en/ip_23_3609 (last visited Nov. 15, 2024).

107 *See The Data Protection Authority and You*, EUROPEAN DATA PROTECTION BOARD, https://www.edpb.europa.eu/sme-data-protection-guide/data-protection-authority-and-you_en (last visited Nov. 15, 2024).

Contractual Clauses (SCCs) and Binding Corporate Rules (BCRs) to ensure that data transferred beyond EU borders receives the same level of protection.[108]

The GDPR is not the whole story. The relatively recent DSA and DMA establish accountability frameworks for online platforms and gatekeepers with obvious data privacy implications.[109] The same can be said of the requirements of rules in relation to transparency, human oversight, and risk management the EU AI Act imposes in relation to high-risk and medium risk systems.[110]

In contrast to the sweeping EU privacy rules and vast supporting bureaucratic apparatus, in the U.S., privacy interests are recognized occasionally, fragmentedly (i.e., in a sector specific way) and with weak regulatory oversight.[111] In the U.S., there is no federal general data privacy law, only sector specific rights and a handful of state laws modeled on aspects of the GDPR.[112] Instead, privacy rights are recognized in isolated contexts, regulated by laws such as the Gramm-Leach-Bliley Act (GLBA) for financial data, the COPPA for children's data, and the

108 *See* Natalie Whitney, *International Data Transfers: Model Contract Clauses vs. Binding Corporate Rules*, GRCI LAW (April 8, 2021), https://www.grcilaw.com/blog/international-data-transfers-model-contract-clauses-vs-binding-corporate-rules. For an explanation of BCRs, *see Binding Corporate Rules*, EUROPEAN COMMISSION, https://commission.europa.eu/law/law-topic/data-protection/international-dimension-data-protection/binding-corporate-rules-bcr_en (last visited Nov. 15, 2024).

109 *See* Peter Chapman, *Advancing Platform Accountability: The Promise and Perils of DSA Risk Assessments*, TECH POLICY. PRESS (Jan 9, 2025), https://www.techpolicy.press/advancing-platform-accountability-the-promise-and-perils-of-dsa-risk-assessments/; *See also The Digital Markets Act: Ensuring Fair and Open Digital Markets*, EUROPEAN COMMISSION, https://commission.europa.eu/strategy-and-policy/priorities-2019-2024/europe-fit-digital-age/digital-markets-act-ensuring-fair-and-open-digital-markets_en (last visited Feb. 9, 2025)

110 *See Understanding the EU AI Act: Requirements and Next Steps*, ISACA (Oct.18, 2024), https://www.isaca.org/resources/white-papers/2024/understanding-the-eu-ai-act/.

111 *See Reforming the U.S. Approach to Data Protection and Privacy*, COUNCIL ON FOREIGN RELATIONS (Jan 2018), https://www.cfr.org/report/reforming-us-approach-data-protection (discussing the limitations of the U.S. sectoral approach, noting that it often prioritizes industry-specific practices over individual privacy rights).

112 *See e.g.,* Cal. Civ. Code § 1798.100 (West 2023); Cal. Bus. & Prof. Code § 22575 (West 2020).

HIPAA for health information.[113] These laws work independently, sometimes leading to gaps and inconsistencies in protections.[114]

In the U.S., enforcement is weak compared to the EU.[115] The Federal Trade Commission (FTC), the primary agency overseeing consumer privacy, lacks the authority to impose significant penalties on first-time offenders.[116] Non-binding guidance on issues like algorithmic transparency allows corporations to selectively adopt best practices without fear of substantial consequences.[117] The one exception to the general small-beer nature of privacy related fines in the U.S. is the $5 billion penalty imposed on Facebook for privacy violations in 2019.[118] However, the circumstances of the fine are telling, Facebook had, according to the FTC, not only repeatedly used deceptive disclosures and settings to undermine users' privacy preferences, thus share users' personal information with third-party apps that were downloaded by the user's Facebook "friends"; it had done so in violation of a 2012 settlement the company had made with the FTC.[119] The FTC also alleged that Facebook took inadequate steps to deal with apps that it knew were violating its platform policies.[120] If the same case

113 *See Gramm-Leach-Bliley Act*, FEDERAL TRADE COMMISSION, https://www.ftc.gov/business-guidance/privacy-security/gramm-leach-bliley-act (last visited Nov. 15, 2024) (protecting financial data); *Children's Online Privacy Protection Rule (COPPA)*, FEDERAL TRADE COMMISSION, https://www.ftc.gov/legal-library/browse/rules/childrens-online-privacy-protection-rule-coppa (last visited Nov.15, 2024); *Health Insurance Portability and Accountability Act of 1996 (HIPAA)*, CENTERS FOR DISEASE CONTROL AND PREVENTION, https://www.cdc.gov/phlp/php/resources/health-insurance-portability-and-accountability-act-of-1996-hipaa.html (Sep. 10, 2024) (establishing federal standards protecting sensitive health information).

114 *See Navigating the Patchwork of Privacy: State Privacy Laws in the Absence of a Federal Framework*, BOSTON UNIVERSITY SCHOOL OF LAW (Aug. 16, 2024), https://sites.bu.edu/dome/2024/08/16/navigating-the-patchwork-of-privacy-state-privacy-laws-in-the-absence-of-a-federal-framework/.

115 *See The FTC is Currently the Primary Privacy Enforcer but its Authority is Limited*, NEW AMERICA, https://www.newamerica.org/oti/reports/enforcing-new-privacy-law/the-ftc-is-currently-the-primary-privacy-enforcer-but-its-authority-is-limited/ (last visited Nov.15, 2024).

116 *See Notices of Penalty Offenses*, FEDERAL TRADE COMMISSION, https://www.ftc.gov/enforcement/penalty-offenses (last visited Nov.16, 2024) (the Commission can seek civil penalties if it proves that the company knew the conduct was unfair or deceptive in violation of the FTC Act and the FTC had already issued a written decision that such conduct is unfair or deceptive).

117 *See Using Artificial Intelligence and Algorithms*, FEDERAL TRADE COMMISSION (April 8, 2020), https://www.ftc.gov/business-guidance/blog/2020/04/using-artificial-intelligence-algorithms (the principles serve more as recommendations rather than enforceable regulations).

118 *See FTC Imposes $5 Billion Penalty and Sweeping New Privacy Restrictions on Facebook*, FEDERAL TRADE COMMISSION (July 24, 2019), https://www.ftc.gov/news-events/news/press-releases/2019/07/ftc-imposes-5-billion-penalty-sweeping-new-privacy-restrictions-facebook

119 *Id.*

120 *Id.*

were brought in the EU today, fines would likely amount to 4% of Meta's global revenue, a number also in the billions.

The prioritization of economic efficiency and national security further weakens privacy protections in the U.S.[121] The Patriot Act, for instance, grants agencies like the NSA and FBI authority to conduct warrantless wiretapping, roving wiretaps, and bulk data collection with minimal oversight.[122] Programs like PRISM and Stellar Wind, conducted in cooperation with major tech companies, provided government agencies with extensive access to stored communications and real-time data from platforms like Google and Microsoft—all in the name of collective defense.[123] In such a framework, privacy is not an absolute right but one that can be subordinated to other priorities.

### 2. *Contrasting origins and aims of EU and U.S. privacy law*

While we do not dispute the foregoing characterization, we contend that its implications have been misinterpreted. The primacy of data privacy rights within the EU legal framework is nominal rather than causal. Or, to put it another way, rights are the means, not the ends. EU privacy rights emerged as a reaction to European historical experience: the recognition that privacy's erosion is an early indicator of totalitarian encroachment. Europeans today are deeply unsettled by threats to the line between public and private life, for good reason. As Hannah Arendt warned, when privacy dissolves, totalitarian control follows closely behind.[124] Once "the mass man" loses that boundary, she says, he's left adrift, without a stable world to anchor him, without a private refuge where meaning can take shape.[125] The intimate life—the passions, the thoughts, the quiet joys—fades into nothingness.[126]

[121] *See* Tajdar Jawaid, *Privacy vs. National Security*, 69 INT'L J. COMPUTER TRENDS & TECH. No. 7 (July 2020) (unpublished manuscript) https://arxiv.org/pdf/2007.12633; Ira S. Rubinstein, Gregory T. Nojeim & Ronald D. Lee, *Systematic Government Access to Personal Data: A Comparative Analysis*, 4 INT'L DATA PRIVACY L. 96 (2014), https://academic.oup.com/idpl/article/4/2/96/734798.

[122] *See The USA PATRIOT Act: Preserving Life and Liberty*, U.S. DEP'T OF JUSTICE, https://www.justice.gov/archive/ll/highlights.htm (last visited Nov. 16, 2024).

[123] *See* T.C. Sottek and Janus Kopfstein, *Everything You Need to Know about PRISM*, THE VERGE (Jul 17, 2013), https://www.theverge.com/2013/7/17/4517480/nsa-spying-prism-surveillance-cheat-sheet; *Secrets, Surveillance, and Scandals: The War on Terror's Unending Impact on Americans' Private Lives*, POGO PROJECT ON GOV'T OVERSIGHT (Sep.7, 2021), https://www.pogo.org/analysis/secrets-surveillance-and-scandals-the-war-on-terrors-unending-impact-on-americans-private-lives.

[124] HANNAH ARENDT, CHAPTER 2 in THE HUMAN CONDITION (2d ed. 1998), https://www.frontdeskapparatus.com/files/arendt.pdf *See also* Henry A. Giroux, *Totalitarian Paranoia in the Post-Orwellian Surveillance State*, 22 CULTURAL STUDIES 108, 108–140 (2014), https://www.tandfonline.com/doi/full/10.1080/09502386.2014.917118 ("For Orwell, the loss of privacy represented a moral and political offense that clearly signaled the nature, power and severity of an emerging totalitarian state").

Most obviously, in Nazi Germany, terror began with the Gestapo's relentless intrusion into private lives. Citizens are urged to report "suspicious" behaviors of neighbors, family members, and colleagues.[127] "Informants"—ordinary Germans—were everywhere, eager to report anything deemed subversive.[128] The pattern continued in post-war Eastern Europe.[129] Conversations held in private homes, cafes, and other public spaces were spied upon and recorded.[130] Today, Alexa, Google Voice, and other smart home technologies perform similar monitoring functions.[131] Smartphones track location data with precision, while wearable devices such as smartwatches record biometric information, from heart rates to sleep patterns, feeding vast databases for analysis.[132] In the face of such pervasive data collection, it seems reasonable to ask if we are staring at the dawn of a technologically enabled totalitarian order in which every action is tracked, every preference scrutinized, and every choice subtly influenced

In this context, the question for European policymakers is not how to make technology more advanced and thus more intrusive, but rather—when technology is already so advanced and intrusive—how to prevent the abuses of the past from happening again. And, more importantly, how can this be achieved in a way that ensures all member states, along with the corporations operating within their borders, uphold these protections? Through this lens,

---

125 *Id. See also José Ortega y Gasset*, INTERNET ENCYCLOPEDIA OF PHILOSOPHY, https://iep.utm.edu/jose-ortega-y-gasset (last visited Nov. 15, 2024). *See also The Public Life*, HAC BARD (Oct. 24, 2011), https://hac.bard.edu/amor-mundi/the-public-life-2011-10-24 ("A life spent entirely in public, in the presence of others, becomes, as we would say, shallow. While it retains its visibility, it loses its quality of rising into sight from some darker ground which must remain hidden if it is not to lose its depth in a very real, non-subjective sense.").

126 *See* Frank Ejby Poulsen, *Arendt on Privacy*, HYPOTHESES (Nov. 22, 2020), https://privacy.hypotheses.org/1371.

127 *See* Sarah Brayne, Sarah Lageson & Karen Levy, *Surveillance Deputies: When Ordinary People Surveil for the State,* 57 LAW & SOC'Y REV. 462 https://onlinelibrary.wiley.com/doi/full/10.1111/lasr.12681.

128 *Id.*

129 *See e.g., Das Leben der Anderen* (Sony Pictures Classics 2006) (directed by Florian Henckel von Donnersmarck).

130 *See generally* Matt Loughlin, *Is the Gestapo Everywhere? The Origins of the Modern Perception of the Secret Police of the Third Reich*, 11 LEGACY 1 (2011), http://opensiuc.lib.siu.edu/legacy/vol11/iss1/5.

131 *See New Research Reveals Alarming Privacy and Security Threats in Smart Homes*, N.Y.U. TANDON SCH. OF ENG'G (Oct. 24, 2023), https://engineering.nyu.edu/news/new-research-reveals-alarming-privacy-and-security-threats-smart-homes.

132 Stuart A. Thompson & Charlie Warzel, *Smartphones Are Spies. Here's Whom They Report To*, N.Y. TIMES (Dec. 20, 2019), https://www.nytimes.com/interactive/2019/12/20/opinion/location-tracking-smartphone-marketing.html.

rights—while explicitly protected in the law and undeniably important for advancing these regulations—are secondary to the more urgent task of containing the relentless advance of technological power. Putting nomenclature to one side, the European Union does not treat privacy rights as intrinsic ends, but rather as functional constraints on both state and corporate actors, aimed at preserving institutional balance. More cynically one could argue that ultimately these rights exist to justify and empower an EU privacy bureaucracy which acts primarily on behalf of the state.

Why did the U.S. respond so differently to the social and technological phenomenon that led Europe to adopt the GDPR? Part of the answer lies in the influence of law and economics as exemplified by Judge Richard Posner who argued that legal decisions should prioritize economic efficiency over the protection of certain fundamental rights when they conflict with economic objectives.[133] Law, Posner asserted, is a tool for maximizing societal wealth and productivity,[134] with the protection of rights hinging on a cost-benefit analysis.[135] Rather than viewing privacy as a fundamental right warranting protection in its own regard, many legal scholars and judges increasingly framed it as an instrumental good—one that individuals leverage to optimize strategic outcomes in commercial, social, and political spheres.[136] Under this logic, privacy is neither sacrosanct nor absolute; instead, its worth is contingent upon its utility within broader economic structures. Privacy, when it impedes market efficiency, makes transparency the more desirable quality, as it offers greater societal benefits.[137] Posner's influence extended far beyond academic debate.[138] The cost-benefit framework he and others advocated reshaped how courts and policymakers approached issues like data privacy,

---

133 *See* Richard Posner, *The Economics of Justice*, 80 MICH. L. REV. 942 (1982) (defending the concept of wealth maximization as an adequate ethical basis for evaluating legal and quasilegal institutions)

134 Margaret S. Hrezo & William E. Hrezo, *Judicial Regulation of the Environment Under Posner's Economic Model of the Law*, 18 J. ECON. ISSUES 1071 (1984).

135 Matthew D. Adler & Eric A. Posner, *Rethinking Cost-Benefit Analysis*, 109 YALE L.J. 165 (1999)

136 Richard A. Posner, *The Right of Privacy*, 12 GA. L. REV. 393, 394 (1978) (Privacy as intermediate goods).

137 Posner, *The Economics of Justice*, *supra* note 141 at 942.

138 Posner contributed significantly—perhaps more than anyone else—to the development of the field of "law and economics," and he influenced torts, contracts, antitrust, and intellectual property law by emphasizing outcomes that maximize social welfare. His books, such as *Economic Analysis of Law* and *The Economics of Justice* provided foundational texts that continue to influence curricula and research agendas in law schools today. *See e.g.,* Jeffrey Lynch Harrison, *Fingerprints: An Impressionistic and Empirical Evaluation of Richard Posner's Impact on Contract Law*, 50 U. PAC. L. REV. 373 (2018) (examining Posner's scholarship's influence on contract law); Hans-Bernd Schäfer & Massimiliano Vatiero, *Introduction: Posner's Economic Analysis of Law at Fifty and the Globalization of Jurisprudence*, 31 HIST. ECON. IDEAS 11 (2023) (discussing the global impacts of law and economics); William F. Baxter, *Posner's Antitrust Law: An Economic Perspective*, 8 BELL J. ECON. 609 (1977) (reviewing Richard A. Posner, *Antitrust Law: An Economic Perspective*) (analyzing Posner's contributions to antitrust law) workplace surveillance, and corporate transparency.[139] This view of privacy has found its way into opinions such as *Mathews v. Eldridge*, where the Supreme Court applied a cost-benefit analysis to weigh administrative efficiency against individual rights,[140] and *Utah v. Strieff*, which balanced the deterrence of police misconduct against the societal costs of excluding unlawfully obtained evidence.[141]

The U.S. response to data privacy issues has favored industry-led regulation and self-regulation over top-down government intervention. American governments have occasionally seen the need for sector specific regulation, such as HIPAA for health data, and GLBA for financial data, but these instruments reflect niche areas of concern, not broad regulatory imperatives. Furthermore, at the urging of Silicon Valley, U.S. law makers have generally resisted adopting broad privacy protections arguing that such regulations slow down technological advancements, increase compliance costs, and limit competition—particularly for smaller companies and startups.[142] Instead, U.S. policymakers have opted for a risk-based, ex-post enforcement model, where regulatory action is taken, if at all, after harm occurs, rather than ex-ante rules that could preemptively restrict innovation. That resistance to regulation has occasionally given way at a state level, with states such as California choosing to enact GDPR-inspired laws.

*

In sum, we agree with the general characterization that the EU has established a significantly higher degree of data privacy regulation than the U.S. But whereas others see these differences as emanating from Europe's regard for fundamental rights an end unto themselves, we see EU law as a pragmatic response to historical concerns about state overreach and surveillance. In this equation, rights are not the end, they are a convenient instrument wielded largely by bureaucracies, not individuals, that constrains state and corporate power. The EU's approach is one of caution and pre-emptive risk mitigation. The U.S. has a very different history and political culture, one in which our less cautious, more fragmented, and ultimately transactional regard for data privacy makes sense.

---

139 *Id.*

140 Andrew Blair-Stanek, *Twombly is the Logical Extension of the Mathews v. Eldridge Test for Discovery*, 62 FLA. L. REV. 1, 11 (2010).

141 *See generally*, *Fourth Amendment-Exclusionary Rule-Deterrence Costs and Benefits-Utah v. Strieff*, 130 HARV. L. REV. 337 (2016).

142 *See e.g.,* Suzanne Smalley, *State Privacy Laws Have Been Crippled by Big Tech, New Report Says*, THE RECORD (Feb 1, 2024), https://therecord.media/state-privacy-laws-big-tech-lobbying-report.

## B. Cybersecurity

In this Section we compare and contrast EU and U.S. Cybersecurity regulation, a field closely related to, but distinct from data privacy. The trans-Atlantic contrast is not as stark as it was for privacy—indeed, there are many areas of complementarity—but this comparative review still demonstrates that EU AI regulation prioritizes stability and risk management over abstract theoretical commitments to a rights-based normative framework. Cybersecurity laws in the EU and U.S. are motivated by shared concerns: the need to protect critical infrastructure, reduce the risks hacking and intrusion, and to build resilience against evolving threats. However, they diverge somewhat in regulatory style. The EU mandates various security measures and security-by-design principles; whereas the U.S. relies on a decentralized, market-driven system, leaving individual sectors to manage risks as they arise.

### *1. EU: preventive risk management*

The EU's cybersecurity strategy seems to follow the logic of the "Immunological Other,"[143] where threats are framed as external, invasive forces that must be identified, excluded, and neutralized to preserve systemic stability. Just as the immune system identifies and neutralizes threats to the body, societies establish mechanisms to protect themselves from perceived dangers by designating certain entities, ideas, or groups as external threats.[144] In this context, cyberattacks, insecure technologies, and systemic vulnerabilities are constructed as invasive forces that must be mitigated to protect the integrity of the system. By eliminating the foreign and abnormal, the vitality of the system itself is reinforced: the fewer external degenerates there are, the more the collective can thrive. In this way, the EU cybersecurity framework treats external cyber threats as the Other to ensure the resilience of critical infrastructure.[145]

[143] Originally, this concept comes from immunology, which studies how organisms defend themselves against external threats like viruses and bacteria. The two key ideas are "self" (the body's own cells and tissues) and "non-self" (foreign invaders). Then, Jacques Derrida and Donna Haraway expanded this biological concept into broader discussions about identity and otherness. Derrida, for instance, explored how the immune system's ability to define "self" and "other" mirrors societal processes of inclusion and exclusion. Haraway, in her essay *The Biopolitics of Postmodern Bodies,* examined how the language of immunology shapes our understanding of the body as a political and cultural entity.

[144] Of course, this analogy isn't merely a functional comparison between the immune system and cybersecurity defenses. It reflects a deeper historical pattern: societies defining themselves by identifying and neutralizing perceived external threats. Jacques Derrida, in *Autoimmunity: Real and Symbolic Suicides*, explores "autoimmunity" as the paradox where a system, in its attempt to protect itself, risks self-destruction. *See Derrida and the Immune System*, ET AL., https://etal.hu/en/archive/terrorism-and-aesthetics-2015/derrida-and-the-immune-system.(last visited Jan. 6, 2025). Defensive mechanisms meant to shield a community can unravel it from within—much like how mass data collection, justified as cybersecurity, corrodes user trust and destabilizes the very infrastructure it claims to secure. Similarly, Roberto Esposito, in *Immunitas: The Protection and Negation of Life* shows how the drive to protect the social body often fosters exclusion, suppressing differences and justifying authoritarian measures under the guise of collective safety. *See generally* ROBERTO ESPOSITO, IMMUNITAS: THE PROTECTION AND NEGATION OF LIFE (Zakiya Hanafi trans., Polity Press 2011). Arguably, Cybersecurity policies that restrict access from specific regions under the pretext of "risk mitigation" show this logic, potentially reinforcing digital divides while offering only a superficial sense of security. While a full exposition of these theories is beyond the length of this paper, we conjecture that in a hyperconnected world, where technology seems to call democracy into question, the pursuit of immunity from external threats calls for a reexamination of past assumptions and a critical interrogation of who or what is deemed a threat.

The EU's cybersecurity approach is predominantly proactive, aiming to prevent,[146] anticipate, and respond to cyber threats within critical infrastructures.[147] Through its "security-by-design" principles, the EU embeds protections directly into the architecture of systems from their earliest stages of development, ensuring that vulnerabilities are addressed before they escalate.[148] Statutes such as the Cyber Resilience Act (CRA) mandate advanced encryption protocols, continuous real-time monitoring, and breach notification requirements to protect sensitive systems.[149] Similarly, the NIS2 Directive widens the scope of cybersecurity measures to include medium and large enterprises, making sure vulnerabilities across member states are identified and mitigated uniformly.[150] The Cybersecurity Act further extends this proactive logic by establishing an EU-wide certification framework for AI and IoT systems, preemptively securing emerging technologies.[151] Together, these measures shield critical systems from external threats, reinforcing systemic stability.[152]

---

[145] *See* Zsolt Bederna & Zoltan Rajnai, *Analysis of the Cybersecurity Ecosystem in the European Union*, 3 INT'L CYBERSECURITY L. REV. 35 (2022) (discussing EU's efforts to establish harmonized cybersecurity standards, highlight the role of directives in addressing vulnerabilities. However, it also notes that the harmonization efforts are not yet complete).

[146] *See also Cybersecurity: How the EU tackles Cyber Threats*, COUNCIL OF THE EUROPEAN UNION, https://www.consilium.europa.eu/en/policies/cybersecurity/ (other preventive mechanism include establishing a network of security operation centers across the EU to monitor and anticipate cyber threats, enabling early detection and response.) (last visited Nov 13, 2024).

[147] *See* Philipp S. Krüger & Jan-Philipp Brauchle, *The European Union, Cybersecurity, and the Financial Sector: A Primer,* CARNEGIE ENDOWMENT FOR INTERNATIONAL PEACE (Mar. 16, 2021), https://carnegieendowment.org/2021/03/16/european-union-cybersecurity-and-financial-sector-primer-pub-84055; *Cybersecurity in the European Union*, COOLEY (Oct. 2, 2024), https://cdp.cooley.com/cybersecurity-in-the-european-union/ (explaining that the EU's directives require organizations to adopt risk-based measures to safeguard critical infrastructure.)

[148] Eldar Haber & Aurelia Tamò-Larrieux, *Privacy and Security by Design: Comparing the EU and Israeli Approaches to Embedding Privacy and Security*, 37 COMP. L. & SEC. REV. 105409 (2020), https://doi.org/10.1016/j.clsr.2020.105409.

[149] *See generally* Amanita Security, *Reflections on Cyber Resilience Act Requirements*, AMANITA SECURITY https://www.amanitasecurity.com/posts/reflections-on-cyber-resilience-act-requirements/ (last visited Jan. 6, 2025).

[150] *Directive on Measures for a High Common Level of Cybersecurity Across the Union (NIS2 Directive)*, EUROPEAN COMMISSION: SHAPING EUROPE'S DIGITAL FUTURE, https://digital-strategy.ec.europa.eu/en/library/nis2-directive (last visited Jan.6, 2025).

Additionally, the EU's framework incorporates reactive elements, shaped by crises that exposed vulnerabilities in digital infrastructure. The original NIS Directive emerged soon after the 2007 cyberattacks on Estonia, where distributed denial-of-service (DDoS) attacks paralyzed the nation's systems.[153] Subsequent updates, like NIS2,[154] responded to the surge in ransomware attacks, including those targeting healthcare systems during the COVID-19 pandemic, which disrupted critical services.[155] The Cyber Resilience Act addressed vulnerabilities exploited by botnet attacks on insecure IoT devices,[156] while the Cybersecurity Act was driven by large-scale incidents like WannaCry and NotPetya, which underscored the fragility of interconnected networks.[157] These measures reflect an evolving strategy, which fortifies defenses by learning from past threats and neutralizing external adversaries—the Other—that exploit systemic vulnerabilities.

To operationalize its cybersecurity strategy, the EU relies heavily on private-sector collaboration, but that collaboration demanded by law, not suggested by non-binding government standards. Under NIS2, private companies must adopt stringent security measures and promptly report breaches.[158] The Digital Operational Resilience Act (DORA) extends this approach by requiring financial institutions to implement rigorous ICT risk frameworks, including stress tests to counter advanced cyberattacks.[159] Similarly, the Cyber Resilience Act mandates that manufacturers embed cybersecurity into their digital products by design and maintain these safeguards throughout the product lifecycle.[160] Like immune cells neutralizing pathogens, these entities continuously monitor, adapt, and respond to emerging cyber threats, preserving the system's resilience.

---

151 *(EU) 2019/881 of the European Parliament and of the Council of 17 April 2019 on ENISA (the European Union Agency for Cybersecurity) and on Information and Communications Technology Cybersecurity Certification and Repealing Regulation (EU) No 526/2013 (Cybersecurity Act)*, 2019 O.J. (L 151) 15, http://data.europa.eu/eli/reg/2019/881/oj.

152 This approach is complemented by mandating advanced encryption protocols to protect sensitive data, and keeping continuous real-time monitoring to detect irregularities as they occur. *See What Does Data Protection "by Design" and "by Default" Mean?*, EUROPEAN COMMISSION, https://commission.europa.eu/law/law-topic/data-protection/reform/rules-business-and-organisations/obligations/what-does-data-protection-design-and-default-mean_en (last visited Feb. 10, 2025) (showing that Art. 25 of the GDPR mandates "data protection by design and by default," requiring organizations to implement appropriate technical measures, including encryption, to protect personal data throughout its processing lifecycle); *see also Understanding the EU Cyber Resilience Act (CRA): An Overview*, CYBELLUM (June 20, 2024) https://cybellum.com/blog/understanding-the-eu-cyber-resilience-act-cra-an-overview/ (explaining that one of CRA objective is to continuously monitor for potential threats).

153 Of course, this is not to claim that the original NIS Directive emerged specifically as a response to the 2007 cyberattacks on Estonia. It simply highlights that the Estonia attacks were a significant early wake up call for European Cybersecurity. *See* Directive 2016/1148 of the European Parliament and of the Council, 2016 *O.J.* (L 194) 1.

154 Directive (EU) 2022/2555 of the European Parliament and of the Council, 2022 *O.J.* (L 333) 80.

155 *NIS-2 Directive: Political Agreement on New Rules on Cybersecurity*, EUROPEAN COMMISSION NEWSROOM (July 13, 2022), https://ec.europa.eu/newsroom/cipr/items/753540.

156 *See 5 Cyber Attacks Caused by IoT Security Vulnerabilities*, GLOBAL CYBERSECURITY ASSOCIATION (last visited Jan 2, 2024) https://globalcybersecurityassociation.com/blog/5-cyber-attacks-caused-by-iot-security-vulnerabilities.

157 *WannaCry Is Not History*, CYBERPEACE INSTITUTE (May 12, 2021), https://cyberpeaceinstitute.org/news/wannacry-is-not-history.

*2. US: Driven by agencies*

The U.S. takes a market-driven approach to cybersecurity, relying on sector-specific regulations and decentralized enforcement rather than a comprehensive, uniform framework.[161] This fragmented model works like a distributed immune system, with sector-specific defenses responding to localized threats and ensuring that no single breach can cripple the entire network.[162] However, this decentralization also creates uneven protections, as less-regulated sectors may lack equally rigorous defenses.

For example, in the finance sector, the Federal Trade Commission (FTC) enforces the Safeguards Rule under the GLBA to address data security and consumer privacy.[163] Financial institutions—such as banks, credit unions, and investment firms—must implement security measures to prevent breaches, including regular testing, third-party vendor oversight, and incident response plans.[164] However, these measures are often reactive, triggered only after breaches occur.

---

158 *NIS2 Requirements: Understand and Prepare for the Upcoming NIS2 Requirements*, NIS2 DIRECTIVE, https://nis2directive.eu/nis2-requirements (last visited Feb. 11, 2025).

159 *Digital Operational Resilience Act (DORA)*, EUROPEAN INSURANCE AND OCCUPATIONAL PENSIONS AUTHORITY, https://www.eiopa.europa.eu/digital-operational-resilience-act-dora_en (last visited Jan. 6, 2025).

160 *See European Cyber Resilience Act*, COBALT, https://www.cobalt.io/blog/european-cyber-resilience-act (last visited Jan. 6, 2025)

161 *See generally* Jeff Kosseff, *Defining Cybersecurity Law*, 103 IOWA L. REV. 985 (2018).

162 *See e.g., Critical Infrastructure Sectors*, AMERICA'S CYBER DEFENSE AGENCY, https://www.cisa.gov/topics/critical-infrastructure-security-and-resilience/critical-infrastructure-sectors (last visited Feb 11, 2025).

163 *Gramm-Leach-Bliley Act*, FEDERAL TRADE COMMISSION, https://www.ftc.gov/business-guidance/privacy-security/gramm-leach-bliley-act (last visited Nov. 13, 2024).

164 *See FTC Safeguards Rule: What Your Business Needs to Know*, FEDERAL TRADE COMMISSION, https://www.ftc.gov/business-guidance/resources/ftc-safeguards-rule-what-your-business-needs-know (last visited Nov. 13, 2024).

In contrast, regulations under HIPAA in the healthcare sector focus on safeguarding patient data through encryption, secure transmission, and strict access controls to protect confidentiality and integrity.[165] HIPAA takes a proactive approach, requiring regular risk assessments and technical safeguards to prevent unauthorized access.[166]

At the same time, broader frameworks like the National Institute of Standards and Technology (NIST) Cybersecurity Framework—widely respected as a global benchmark for managing cybersecurity risks—remain voluntary.[167] Adoption depends on market incentives and the willingness of organizations to follow best practices.[168] While sectors like defense and critical infrastructure increasingly embrace NIST to enhance resilience, industries without specific regulatory mandates still lag behind.[169]

To address such gaps in the sectoral regulations in the U.S., some critics have argued for a unified, EU-style framework.[170] They suggest that comprehensive federal laws, similar to the EU's NIS2 Directive or GDPR, could streamline governance and ensure consistent protections for critical infrastructure and personal data.[171] We are agnostic on the merits of such proposals, but it would be wrong to overlook the benefits of a decentralized system rooted in pluralism and competition.[172]

---

165 *See HIPAA Encryption Requirements*, HIPAA JOURNAL, https://www.hipaajournal.com/hipaa-encryption-requirements (last visited Jan. 6, 2025); *See also Security Standards: Technical Safeguards*, HHS.GOV, https://www.hhs.gov/sites/default/files/ocr/privacy/hipaa/administrative/securityrule/techsafeguards.pdf (last visited Jan. 6, 2025).

166 *Id.* The key differences between HIPPA and GLBA is that HIPAA focuses on medical records and treatment plans while GLBA focuses on NPI; HIPAA mandates controls directly tied to healthcare ecosystems while GLBA requires protocols relevant to financial services, such as protecting consumer financial transactions. And HIPAA has explicit breach notification requirements while GLBA doesn't specifically mandate breach notifications.

167 *NIST Cybersecurity Framework*, FTC.GOV, https://www.ftc.gov/business-guidance/small-businesses/cybersecurity/nist-framework (last visited Jan. 6, 2025).

168 *Id.*

169 *Critical Infrastructure Protection: Additional Actions Are Essential for Assessing Cybersecurity Framework Adoption*, U.S. GOV'T ACCOUNTABILITY OFFICE (Feb. 15, 2018), https://www.gao.gov/products/gao-18-211. (reporting that most of the 16 critical infrastructure sectors have taken steps to facilitate the adoption of the NIST CSF).

170 *See* Radanliev, *Review and Comparison of US, EU, and UK Regulations on Cyber Risk/Security of the Current Blockchain Technologies: Viewpoint from 2023*, 17 BLOCKCHAIN TECH. REV. 105 (2023), https://link.springer.com/article/10.1007/s12626-023-00139-x (provides a comparative review of US, EU, and UK regulatory approaches, highlighting gaps and differences between them; implicitly suggesting that the US could improve its cybersecurity posture by adopting more cohesive regulatory approaches).

171 *See A Guide to U.S. Cybersecurity Laws and Compliance*, NRI SECURE (Dec.5, 2024), https://www.nri-secure.com/blog/us-cybersecurity-laws-compliance; *See also* Moira Warburton, *US Lawmakers Push for Federal Data Privacy Law; Tech industry and Critics are Wary*, REUTERS (June 26, 2024), https://www.reuters.com/world/us/federal-data-privacy-laws-gain-support-us-congress-critics-remain-2024-06-26.

A decentralized approach to cybersecurity has allowed industries to address specific risks with tailored solutions. In the energy sector, for instance, the North American Electric Reliability Corporation (NERC) developed its Critical Infrastructure Protection (CIP) standards to address the growing threat of cyberattacks on the power grid.[173] These standards, introduced after the 2003 Northeast Blackout exposed critical vulnerabilities, established mandatory requirements for securing industrial control systems, and significantly reduced the risk of large-scale blackouts caused by cyber intrusions.[174] Yet, CIP standards primarily apply to bulk power systems.[175] To suggest that the energy sector adopt the same standards as healthcare would require a complete overhaul of regulatory frameworks and responsibilities—an almost unworkable task given the distinct operational and risk profiles of each sector.

Once again, while the goals of cybersecurity may align across the EU and the U.S., their approaches reflect political differences. The EU's strategy treats cyber threats as invasive forces to be identified and neutralized across the system, while the U.S. addresses localized threats independently. There is nothing in EU cybersecurity regulation that suggests a rights-based normative commitment. The EU's approach is manifestly about precaution and risk-regulation. Europe might be right to be more prescriptive in this field and rely less on soft law, but that is an argument that should be made on the merits of precautionary regulation, not the presumption that the European approach is intrinsically rights-regarding.

## C. Healthcare and Technology

The regulation of AI in healthcare is another area that vindicates our thesis that EU regulation reflects a long tradition of managing risk and preserving stability, rather than a commitment to fundamental rights. Once again, a comparison with the U.S. is illuminating, although there are many areas of convergence. In general, the EU's healthcare technology regulations focus on rigorous pre-market evaluations, strong human oversight, comprehensive data privacy protections, and proactive post-market surveillance. Although these features are not entirely alien to the U.S. health law landscape, in general contrast, the U.S. prioritizes speed, innovation, and interoperability, often at the cost of thorough risk assessments and data security. We see this contrast in four areas: the EU has a stricter regulatory pathway for medical technology including AI; the EU is more insistent that humans remain "in the loop" for medical decisions; the EU's more stringent approach to medical data; and the EU's more expansive and proactive post-market surveillance framework.[176]

[172] *See generally* Lior Jacob Strahilevitz, *Toward a Positive Theory of Privacy Law*, 126 HARV. L. REV. 2010 (2013) (noting that U.S. lacks a comprehensive privacy framework, thereby has a sectoral approach that addresses privacy concerns within specific industries rather than through overarching legislation); *See also* Daniel J. Solove & Chris Jay Hoofnagle, *A Model Regime of Privacy Protection*, 2006 U. ILL. L. REV. 357 (2006) (criticizing the US sectoral approach to privacy, which regulates specific industries while leaving others unregulated, leading to gaps and inconsistencies in privacy protection).

[173] *The Evolution of NERC CIP Compliance: Safeguarding the Power grid*, NETWORK PERCEPTION (July 25, 2023), https://www.network-perception.com/blog/the-evolution-of-nerc-cip-compliance.

[174] *Id.*

[175] *Id.*

To begin with our first point of comparison, the EU insists on much more rigorous and time-consuming approval process for medical devices and healthcare technologies than the U.S. In the EU, medical devices reach patients after an exhaustive process of pre-market safeguards.[177] Devices undergo rigorous clinical trials, meet safety and performance standards, and pass detailed risk assessments under the Medical Device Regulation (MDR) and In Vitro Diagnostic Medical Device Regulation (IVDR).[178] This process, which typically takes up to 18 months, makes sure that only devices meeting the high safety and effectiveness standards are allowed on the market.[179] The U.S. takes a different approach, arguably prioritizing speed over thoroughness. With the Food and Drug Administration (FDA's) expedited 510(k) clearance and De Novo pathways, devices can enter the market in just three to six months.[180] The AI diagnostic tool IDx-DR, for example, was cleared in two months.[181] IDx-DR is diagnostic tool designed to detect diabetic retinopathy—a complication of diabetes that can lead to blindness.[182] It was the first fully autonomous AI system approved by the U.S. FDA in 2018 for making medical decisions without requiring a specialist's review. The system has obvious advantages and potential cost savings, but some have expressed concern that, compared to the broader range of conditions a human physician might observe, the AI tool "may give PCPs and patients a false sense of security about the totality of their ocular status."[183] The strictness of EU regulation of AI in health is also apparent in post-market actions. When companies fail to comply with safety protocols, the EU enforces its regulations with strict consequences. Penalties range from complete market withdrawal and sales bans until all issues are resolved to substantial fines.[184] These measures ensure immediate corrective action and uphold patient safety. In contrast, the U.S. typically imposes lighter penalties, such as financial fines or CMS reimbursement denials.[185] While the FDA can issue warning letters, injunctions, or seize non-compliant products, the initial consequences are far less severe compared to the EU's enforcement.[186]

---

[176] We also note a fifth contrast in passing: state investment in medical AI. The U.S. is investing significantly more financial capital in developing AI solutions to medical problems. Programs like the NIH's Bridge2AI initiative, for instance, have dedicated $130 million to advancing AI in biomedical research. *See NIH launches Bridge2AI Program to Expand the Use of Artificial Intelligence in Biomedical and Behavioral Research*, NIH (Sep 13, 2022), https://www.nih.gov/news-events/news-releases/nih-launches-bridge2ai-program-expand-use-artificial-intelligence-biomedical-behavioral-research. In 2022 alone, the NIH funded over 500 AI and machine learning projects. *See Artificial Intelligence*, U.S. NATIONAL SCIENCE FOUNDATION, https://new.nsf.gov/focus-areas/artificial-intelligence (last visited Nov. 14, 2024). Similarly, the Department of Health and Human Services (HHS) has allocated nearly $129 million for AI and AI-related purchases over the past five years. *See HHS has Spent $129 million on AI Purchases in the Past 5 Years, Data Shows*, POLITICOPRO (Sep. 13, 2024), https://subscriber.politicopro.com/article/2024/09/hhs-has-spent-129-million-on-ai-purchases-in-the-past-5-years-data-shows-00178640. No single EU agency has invested this much in the same cause.

[177] Council Regulation 2017/745, Art. 61, 2017 *O.J.* (L 117) 1 (EU) (hereafter, "MDR") (specifying that manufacturers must conduct a clinical evaluation to verify the device's conformity with safety and performance requirements)

[178] *See* MDR at Art.61 (detailing the requirements for clinical evaluations); *Id* at Art.10(2) (requiring manufacturing to implement a risk management system throughout the device's lifecycle). *See also* Regulation (EU) 2017/746 of the European Parliament and of the Council, art 68, 2017 O.J. (L 117) 176.

[179] *Bottlenecks, Timelines, and Complexity: Overcoming EU MDR Challenges*, ARROTEK, https://arrotek.com/bottlenecks-timelines-and-complexity-overcoming-eu-mdr-challenges (last visited Nov 14, 2024).

Our second point of comparison concerns the presence or absence of humans "in the loop" for medical decisions in the EU and the U.S. In the EU, medical devices don't make decisions alone. Even the most advanced AI-assisted systems require a human to validate their outputs.[187] Doctors, not algorithms, hold the authority to diagnose or treat, thereby making sure that clinical expertise remains central.[188] In radiology, for example, AI tools may flag abnormalities, but the final word comes from the radiologist.[189] One reason for this cautious approach is that algorithms are only as good as the data they're fed. A machine trained on incomplete or biased datasets—common in fields like dermatology or cardiology—can misdiagnose groups such as women or people with darker skin.[190] By insisting on human oversight, the EU limits these risks. Granted, human oversight alone won't solve all the problems with algorithmic decision-making. Evidence shows that people often fail to catch an algorithm's flaws.[191] Automation bias is an example—faced with machine outputs, humans tend to trust them, even when they're clearly wrong.[192] Human oversight in these cases becomes little more than rubber-stamping, where the human reviewer blindly approves whatever the algorithm suggests without real scrutiny.[193] Still, the EU's insistence on human oversight slows the slide into a world where machines operate unchecked.

---

180 *Breakthrough Devices Program*, U.S. FOOD & DRUG ADMIN. (Nov.7, 2024), https://www.fda.gov/medical-devices/how-study-and-market-your-device/breakthrough-devices-program.

181 *See* Keng Jin Lee, *AI device for Detecting Diabetic Retinopathy Earns Swift FDA Approval*, AMERICAN ACADEMY OF OPHTHALMOLOGY (Apr 12, 2018), https://www.aao.org/education/headline/first-ai-screen-diabetic-retinopathy-approved-by-f.

182 Michael F. Chiang, *Artificial Intelligence Getting Smarter! Innovations from the Vision Field*, NIH DIRECTOR'S BLOG (Feb 8, 2022), https://directorsblog.nih.gov/tag/idx-dr/.

183 *See* A. Paul Chous, *Pros and cons of using an AI-based diagnosis for diabetic retinopathy.* OPTOMETRY TIMES (Aug 1 2018), https://www.optometrytimes.com/view/pros-and-cons-using-ai-based-diagnosis-diabetic-retinopathy

184 *See Consequences of Non-Compliance*, OBELIS GROUP (June 10, 2019), https://www.obelis.net/news/consequences-of-non-compliance/.

185 45 C.F.R. § 160, subpt. D (2024) (non-compliance imposes civil money penalties). The Centers for Medicare & Medicaid Services (CMS) is the U.S. government agency that oversees Medicare and Medicaid, which provide health coverage to millions of Americans. Without CMS reimbursement, most medical devices are unprofitable, to say the least.

186 E.g., the Essure birth control device, manufactured by Bayer, caused several complications in women, including uterine perforation, migration of the implant, and hair loss. In the U.S., despite these concerns, the device was not immediately pulled from the market. The FDA issued a black box warning in 2016 and imposed sales restrictions. However, despite additional post-market surveillance, Essure remained available in the U.S. until Bayer voluntarily withdrew it from the market in 2018 due to declining sales. In contrast, in 2017, Essure was removed from the EU market entirely after Bayer was unable to renew the device's CE marking, failing to meet stricter safety requirements under the Medical Device Regulation (MDR).

In the U.S., where speed and innovation are prioritized, AI has become the "standard of care."[194] Algorithms lead, and doctors follow.[195] Take Viz.ai, for example. Its stroke detection software scans for large vessel occlusions, flags critical cases, and sends alerts directly to physicians.[196] While human review is still part of the process, if clinicians can't meaningfully challenge the algorithm's findings, its recommendations will ultimately take precedence over a more thoughtful, clinician-led evaluation. As mentioned above, IDx-DR takes this even further. Designed to detect diabetic retinopathy, it operates without a specialist's oversight.[197] The logic seems clear: faster diagnoses, less burden on overworked doctors.[198] The doctor becomes a technician following the algorithm's lead or is replaced by a technician entirely. The promise here is efficiency and scalability, but the tradeoffs in terms of oversight and accountability are uncertain.

---

187 *Medical Artificial Intelligence: The European Legal Perspective*, COMMUNICATIONS ACM (Nov 1, 2021), https://cacm.acm.org/opinion/medical-artificial-intelligence/.

188 *Id.*

189 *See* Elizabeth Short, *AI is Not Ready to Replace Radiologists Interpreting Chest X-Rays*, MEDPAGE TODAY (September 26, 2023), https://www.medpagetoday.com/radiology/diagnosticradiology/106508.

190 *See* Adriana Krasniansky, *Understanding Racial Bias in Medical AI Training Data*, ROCK HEALTH BLOG (Oct. 29, 2019), https://rockhealth.com/insights/understanding-racial-bias-in-medical-ai-training-data/.

191 *See* Ben Green, *The Flaws of Policies Requiring Human Oversight of Government Algorithms*, 45 COMPUT. L. & SEC. REV. 105681 (2022).

192 Automation bias describes the tendency to trust in the outputs of automated decision-making systems, even to the point of ignoring contradictory information. Linda J. Skitka, Kathleen L. Mosier & Mark Burdick, *Does Automation Bias Decision-Making?*, 51 INT'L J. HUM.-COMPUT. STUD. 991 (1999). For a classic text in the field of human factors and automation, discussing the unintended consequences of increasing automation in complex systems, including, what we would now term "automation bias", see Lisanne Bainbridge, *Ironies of Automation*, 19 AUTOMATICA 775 (1983), https://ckrybus.com/static/papers/Bainbridge_1983_Automatica.pdf.

193 Rebecca Crootof, Margot E. Kaminski & W. Nicholson Price II, *Humans in the Loop*, 76 VAND. L. REV. 429, 442 (2023).

194 A. Michael Froomkin, Ian Kerr & Joelle Pineau, *When AIs Outperform Doctors: Confronting the Challenges of a Tort-Induced Over-Reliance on Machine Learning*, 61 ARIZ. L. REV. 33, 72–73 (2019).

195 *See* Cestonaro, Clara, et al., *Defining Medical Liability When Artificial Intelligence Is Applied on Diagnostic Algorithms: A Systematic Review*, 10 FRONTIERS IN MED. 1305756 (2023), https://pmc.ncbi.nlm.nih.gov/articles/PMC10711067/ (If AI algorithms will be integrated into radiology standard of care, deviations from AI readout may indeed prompt liability.)

Our third point of comparison brings us back to data protection. The EU enforces strict data privacy protections under the GDPR, which, as described above, establishes centralized and uniform rules for handling sensitive patient data.[199] These rules are supported by heavy penalties designed to ensure compliance isn't optional and that careless or exploitative data practices are discouraged.[200] The U.S., in contrast, takes a far more permissive approach. Under HIPAA, data protection rules only apply to "covered entities" such as healthcare providers and insurers.[201] As a result, technology firms and app developers, who often process large volumes of sensitive patient data, operate largely outside HIPAA's reach. This gap allows

---

196 *Viz.ai Artificial Intelligence Stroke Software Helping Doctors Win Race Against Time*, RADIOLOGY BUSINESS, https://radiologybusiness.com/sponsored/22221/vizai/topics/artificial-intelligence/vizai-artificial-intelligence-stroke-software (last visited Jan 6, 2025).

197 A. Michael Froomkin, Ian Kerr & Joelle Pineau, *When AIs Outperform Doctors: Confronting the Challenges of a Tort-Induced Over-Reliance on Machine Learning*, 61 ARIZ. L. REV. 33, 44 (2019).

198 But whether the result would actually be more accurate, and the human doctor will be more skillful in diagnosis is another story. *See* Marina Chugunova & Daniela Sele, *We and It: An Interdisciplinary Review of the Experimental Evidence on How Humans Interact with Machines*, 99 J. BEHAV. & EXPERIMENTAL ECON. 1, 2–3 (2022) (reviewing human-computer interactions); Christoph Engel & Nina Grgić-Hlača, *Machine Advice with a Warning About Machine Limitations: Experimentally Testing the Solution Mandated by the Wisconsin Supreme Court*, 13 J. LEGAL ANALYSIS 284, 286 (2021) (experimentally evaluating the effects of algorithmic accuracy warnings and finding limited effects). Sometimes humans don't oversee decisions, *see* Michael Veale & Lilian Edwards, *Clarity, Surprises, and Further Questions in the Article 29 Working Party Draft Guidance on Automated Decision-Making and Profiling*, 34 COMPUT. L. & SEC. REV. 398, 400 (2018). Sometimes humans over-rely on machines. Raja Parasuraman & Dietrich H. Manzey, *Complacency and Bias in Human Use of Automation: An Attentional Integration*, 52 HUM. FACTORS 381, 390–98 (2010); Danielle Keats Citron, *Technological Due Process*, 85 WASH. U. L. REV. 1249, 1271–72 (2008). Sometimes using much AI results in humans' skill fade. *See* Meg L. Jones, *The Ironies of Automation Law: Tying Policy Knots with Fair Automation Practices Principles*, 18 VAND. J. ENT. & TECH. LAW 77, 112 (2020) ("Automation leads to the deterioration of human operator skill, which needs to be more sophisticated to deal with novel and unique situations."); see also Bainbridge, *supra* note 199.

199 *Supra* Part II.A

200 *See What are the GDPR Fines?,* GDPR. EU https://gdpr.eu/fines/ (last visited Jan. 7, 2025)

201 *See Are You a Covered Entity?*, CMS.GOV, https://www.cms.gov/priorities/key-initiatives/burden-reduction/administrative-simplification/hipaa/covered-entities (last visited Nov. 14, 2024) medical technology providers to collect, analyze, and even monetize patient data with minimal oversight.[202]

The Office of the National Coordinator for Health Information Technology (ONC) further demonstrates the U.S. approach. By prioritizing interoperability—the seamless sharing of electronic health records (EHRs) between systems — ONC aims to improve care coordination and streamline healthcare delivery.[203] However, this emphasis on data flow often overlooks downstream risks.[204] Once data leaves its original source, it can be repurposed, shared, or even sold with little regulatory intervention.

The divergence in data processing norms between the European Union and the United States can be contextualized within the broader debate over privacy as a marketable asset. In the United States, where patient data is leveraged to enhance healthcare efficiency, optimize resource allocation, and drive technological advancement, it is unsurprising that privacy protections are comparatively weaker than in the EU. This distinction was underscored in *Dinerstein v. Google*,[205] in which a federal appellate court rejected a claim that a hospital unlawfully compromised patient privacy by sharing anonymized electronic health record data with Google for artificial intelligence research. The court found that the plaintiff failed to establish standing, reasoning that the alleged privacy violation did not constitute a concrete injury.[206]

Finally, we consider the European Union's post-market surveillance framework, which is both more expansive and proactive in its approach. Post-market surveillance entails the ongoing assessment of medical devices' safety and efficacy following regulatory approval and clinical deployment.[207] This is the stage where hidden risks—those not apparent during pre-market evaluations—emerge. In the EU, post-market surveillance is built on constant oversight, with systems designed to anticipate and address risks before they escalate.[208]

---

[202] *Out of Control: Dozens of Telehealth Startups Sent Sensitive Health Information to Big Tech Companies*, THE MARKUP (Dec. 13, 2022), https://themarkup.org/pixel-hunt/2022/12/13/out-of-control-dozens-of-telehealth-startups-sent-sensitive-health-information-to-big-tech-companies.

[203] *See* Tayla Holman, *ONC (Office of the National Coordinator for Health Information Technology)*, TECHTARGET https://www.techtarget.com/searchhealthit/definition/ONC (last visited Jan.7, 2025); *See generally Procuring Interoperability: Achieving High-Quality, Connected, and Person-Centered Care*, NATIONAL ACADEMY OF MEDICINE (2018), https://doi.org/10.17226/27114 (underscoring how interoperability ensures that healthcare providers can access and share comprehensive patient data, demonstrating ONC's critical role in standardizing data movement).

[204] *See e.g.,* Anura S. Fernando, *Chapter 4: Interoperability Risks and Health Informatics,* in DIABETES DIGITAL HEALTH AND TELEHEALTH 43, 43–50 (2022), https://www.sciencedirect.com/science/article/pii/B9780323905572000133.

[205] 73 F.4th 502, 510 (7th Cir. 2023).

[206] *Id.*

The EU's regulatory framework for medical devices is built upon a system of Notified Bodies—organizations appointed by individual member states and subject to oversight by the European Commission.[209] These bodies do not merely facilitate market entry; they remain engaged throughout a device's entire lifespan.[210] Annual inspections verify continued compliance, while clinical evaluations provide a mechanism for reassessing safety and efficacy.[211] When risks materialize, National Competent Authorities (NCAs) step in.[212] These bodies serve as the enforcement arm, swiftly investigating potential defects, imposing remedial obligations, and, in critical cases, ordering market withdrawal.[213] The purpose of this layered approach is to ensure that ideally, no risk slips through the cracks.

The U.S., by contrast, relies on a far leaner and more reactive system. Post-market oversight is primarily handled by the FDA,[214] which depends on programs like the Medical Device Reporting (MDR) system and MedWatch.[215] These programs rely on voluntary reporting,

---

[207] For an explanation of post-market surveillance, *See Guidance for Post-Market Surveillance and Market Surveillance of Medical Devices, including In Vitro Diagnostics*, WORLD HEALTH ORGANIZATION (June 20, 2021), https://www.who.int/publications/i/item/9789240015319.

[208] Oversight over their products is constant because MDR. Art.83 mandates that manufacturers implement a PMS system proportionate to the risk class and appropriate for the device type; Art. 84 requires manufacturers to develop a Post-Market Surveillance Plan. The Plan should define the methods and processes for proactively collecting and evaluating data from post-market activities.

[209] *See also Notified Bodies for Medical Devices*, EUROPEAN COMMISSION https://health.ec.europa.eu/medical-devices-topics-interest/notified-bodies-medical-devices_en (last visited Nov. 14, 2024).

[210] *See* Josep Pane et al*., EU Postmarket Surveillance Plans for Medical Devices, 28* PHARMACOEPIDEMIOL. DRUG SAF. 1155, 1155–65 (2019), https://doi.org/10.1002/pds.4859.

[211] *Role of Notified Bodies*, GMED https://lne-gmed.com/notified-bodies-role (last visited Nov. 14, 2024).

[212] *See National Competent Authorities (Humans*), EUROPEAN MEDICINES AGENCY, https://www.ema.europa.eu/en/partners-networks/eu-partners/eu-member-states/national-competent-authorities-human (last visited Jan. 7, 2025).

[213] *See* European Commission, *Guidelines on Medical Devices: Vigilance System*, at 27-32, MEDDEV 2.12-1 REV. 8 (May 2019) https://www.medical-device-regulation.eu/wp-content/uploads/2019/05/2_-12-1_rev8_en.pdf. (manufacturers are required to report serious incidents and field safety corrective actions to the relevant NCAs; NCAs can also take corrective actions such as updates to user instructions).

[214] *Understanding CDER's Postmarket Safety Surveillance Programs and Public Data*, U.S. FOOD & DRUG ADMIN. (April 3, 2024), https://www.fda.gov/drugs/cder-conversations/understanding-cders-postmarket-safety-surveillance-programs-and-public-data.

[215] *Overview of Device Regulation*, U.S. FOOD & DRUG ADMIN. (April 3, 2024), https://www.fda.gov/medical-devices/device-advice-comprehensive-regulatory-assistance/overview-device-regulation (explaining The MDR regulation is a mechanism for FDA and manufacturers to identify and monitor significant adverse events

which means risks are often recognized only after they have caused harm.[216] While the FDA has the authority to mandate post-approval studies for certain high-risk devices,[217] it lacks the EU's robust, multi-agency infrastructure and its proactive, lifecycle-based auditing.

*

As in our previous discussions, we do not mean to suggest that the U.S. approach regulating the use of AI in healthcare is better or worse than in the EU. This appears to be one of many areas in which there are no right answers, only tradeoffs between different priorities. Our point remains that at the intersection of AI and healthcare, EU AI regulation is far less a reflection of a rights-based normative vision than it is a pragmatic institutional response focused on preserving systemic stability and mitigating technological risks within a complex regulatory landscape.

## D. Labor and Employment

We now turn to the regulation of AI in relation to labor and employment. Both the EU and the U.S. recognize the growing risks of AI in employment.[218] Their responses, however, show different priorities. The EU, shaped by a history of labor unrest and systemic exploitation, imposes strict protections to ensure transparency and fairness.[219] In the EU, employers must involving medical devices. The goals of the regulation are to detect and correct problems in a timely manner.) *See MedWatch Forms for FDA Safety Reporting*, U.S. FOOD & DRUG ADMIN. (Feb. 8, 2024), https://www.fda.gov/safety/medical-product-safety-information/medwatch-forms-fda-safety-reporting.

[216] *See e.g., MedWatch Forms for FDA Safety Reporting*, U.S. FOOD & DRUG, https://www.fda.gov/safety/medical-product-safety-information/medwatch-forms-fda-safety-reporting (last visited Jan 7, 2025). However, it needs to be noted that reporting is mandatory from manufacturers and importers.

[217] For high-risk devices, such as implantable cardiac devices or automated insulin delivery systems, the FDA may require post-approval studies or Section 522 postmarket surveillance studies to collect long-term safety data. *See Postmarketing Surveillance Programs*, U.S. FOOD & DRUG ADMIN. (April 2, 2020), https://www.fda.gov/drugs/surveillance/postmarketing-surveillance-programs.

[218] *See Addressing AI Risks in the Workplace: Workers and Algorithms*, EUROPEAN PARLIAMENT (June , 2024), https://www.europarl.europa.eu/RegData/etudes/BRIE/2024/762323/EPRS_BRI(2024)762323_EN.pdf; *Addressing AI Risks in the Workplace: Workers and algorithms*, THINK TANK EUROPEAN PARLIAMENT (June 3, 2024), https://www.europarl.europa.eu/thinktank/en/document/EPRS_BRI(2024)762323. For the discussions of harms in the US, *See Artificial Intelligence And Worker Well-being: Principles And Best Practices For Developers And Employers*, U.S. DEPARTMENT OF LABOR, https://www.dol.gov/general/ai-principles. (last visited Nov.15, 2024); *EEOC Hearing Explores Potential Benefits and Harms of Artificial Intelligence and other Automated Systems in Employment Decisions*, U.S. EQUAL EMPLOYMENT OPPORTUNITY COMMISSION (Jan 31, 2023), https://www.eeoc.gov/newsroom/eeoc-hearing-explores-potential-benefits-and-harms-artificial-intelligence-and-other.

[219] *See e.g., How The EU Improves Workers' Rights and Working Conditions*, EUROPEAN PARLIAMENT (May 14, 2019), https://www.europarl.europa.eu/topics/en/article/20190506STO44344/how-the-eu-improves-workers-rights-and-working-conditions.

disclose AI's role, secure explicit consent, and remain accountable for its impacts. The U.S., by contrast, relies on corporate self-regulation. It trusts that businesses—driven by innovation and market demands—are best positioned to address AI's workplace challenges.[220]

The EU's labor protections go beyond safeguarding individual rights; they aim to prevent the destabilizing effects of unregulated AI misuse.[221] Regulation in the EU is motivated by concerns that AI systems used to track productivity, monitor communications, or analyze biometric data can easily become tools for employer control and exploitation.[222] These technologies, critics warn, risk deepening inequalities, undermining worker autonomy, and fostering disempowerment and distrust.[223] To counter this, EU directives require transparency, obligating employers to disclose when and how these systems evaluate workers.[224] They are designed to preserve workplace stability by addressing the historical risks of unchecked technological power.

This regulatory stance might be contextualized in Europe's historical awareness of how labor exploitation can lead to upheaval. During the Industrial Revolution, technological advancements enabled employers to impose dehumanizing conditions on workers.[225] Friedrich Engels, in *The Condition of the Working Class in England*, described how unchecked industrial progress concentrated wealth and power at workers' expense.[226] Together with Karl Marx, he warned in *The German Ideology* that such systemic exploitation would inevitably provoke revolt: "The proletarians cannot emancipate themselves without at the same time abolishing their own previous mode of existence."[227]

---

220 *See e.g.,* Melissa Heikkila, *AI companies promised to self-regulate one year ago. What's changed?*, MIT TECHNOLOGY REVIEW (July 22, 2024), https://www.technologyreview.com/2024/07/22/1095193/ai-companies-promised-the-white-house-to-self-regulate-one-year-ago-whats-changed (arguing that traditionally, the US has been loath to regulate its tech giants, instead relying on them to regulate themselves).

221 *See Commission Welcomes Political Agreement on Improving Working Conditions in Platform Work*, EUROPEAN COMMISSION (Dec 12, 2023), https://ec.europa.eu/commission/presscorner/detail/en/ip_23_6586.

222 *See* Chiara Litardi, *Employee Monitoring: A Moving Target for Regulation*, EUROFOUND (July 15, 2024), https://www.eurofound.europa.eu/en/resources/article/2024/employee-monitoring-moving-target-regulation.; *See also Data Subjects, Digital Surveillance, AI and the Future of Work*, EUROPEAN PARLIAMENT (2020), https://www.europarl.europa.eu/RegData/etudes/STUD/2020/656305/EPRS_STU%282020%29656305_EN.pdf?ref=legal-digital (discussing GDPR's requirements on employers for processing workers' personal data, EU Charter of Fundamental Rights' guarantee for the workers to challenge intrusive surveillance practices, the existing labor laws, such as the Working Time Directive and the Framework Directive on Health and Safety at Work to address stress, over-monitoring, and the psychosocial effects of surveillance).

223 *See AI @ Work: Human Empowerment or Disempowerment?*, INTRODUCTION TO DIGITAL HUMANISM 175, 175–96 (Springer Nature 2023), https://link.springer.com/chapter/10.1007/978-3-031-45304-5_12; *See also generally,* IFEOMA AJUNWA, THE QUANTIFIED WORKER: LAW AND TECHNOLOGY IN THE MODERN WORKPLACE (Cambridge Univ. Press 2023).

224 *See e.g., What Does the EU AI Act Mean for Employers*, CLIFFORD CHANCE, https://www.cliffordchance.com/content/dam/cliffordchance/briefings/2024/08/what-does-the-eu-ai-act-mean-for-employers.pdf (last visited Nov. 15, 2024).

225 *See* FRIEDRICH ENGELS, THE CONDITION OF THE WORKING CLASS IN ENGLAND (1845), https://www.marxists.org/archive/marx/works/download/pdf/condition-working-class-england.pdf.

While today's risks differ in form, they echo similar systemic challenges. AI does not create immediate physical harm, as industrial machinery once did, but it poses subtler dangers: the erosion of agency, the invasion of privacy, and decisions driven by algorithms.[228] Algorithmic opacity and discriminatory data patterns could marginalize workers, while productivity metrics might encourage relentless surveillance.[229] If mismanaged, such conditions could foster a new form of discontent, echoing previous cycles of resistance to oppressive systems.[230]

The EU's framework addresses these modern risks head-on. By enforcing transparency, mandating fairness, and requiring human oversight, it, supposedly, aims to ensure that AI is used to support workers rather than control them.[231] These measures prevent AI from quietly reinforcing employer exploitation, keeping workplace power in check.

The U.S., by contrast, has historically taken a more permissive approach to labor regulation.[232] Socialist and communist movements—such as Eugene V. Debs' leadership, the 1919 Seattle

---

226 *Id.*

227 *See generally* KARL MARX & FRIEDRICH ENGELS, THE GERMAN IDEOLOGY (1845), "The Ruling Class and the Ruling Ideals," https://web.mit.edu/uricchio/Public/Documents/Marx%20&%20Engels.pdf. *Id* at Part I, https://www.marxists.org/archive/marx/works/1845/german-ideology/ch01a.htm (arguing that morality isn't an abstract, immutable set of principles but a reflection of the material and economic conditions of society. They also dismiss the notion of religion as a divine or eternal truth, viewing it instead as a product of human activity and material circumstances, proposing that individuals are shaped by their social relations and the material conditions in which they live.)

228 *See* Somendra Narayan, *AI and the Future of Human Agency: Are We Outsourcing Decision-Making or Evolving with Machines?*, MEDIUM (Oct.18, 2024) https://medium.com/@narayan.somendra/ai-and-the-future-of-human-agency-are-we-outsourcing-decision-making-or-evolving-with-machines-78da6ba4475f.

229 *See How the Use of AI Impacts Marginalized Populations in Child Welfare*, NC STATE CENTER FOR FAMILY AND COMMUNITY ENGAGEMENT (December 2, 2024), https://cface.chass.ncsu.edu/news/2024/12/02/how-the-use-of-ai-impacts-marginalized-populations-in-child-welfare. *See also* SHOSHANA ZUBOFF, THE AGE OF SURVEILLANCE CAPITALISM: THE FIGHT FOR A HUMAN FUTURE AT THE NEW FRONTIER OF POWER (PublicAffairs 2019).

230 *See* Sahajveer Baweja & Swapnil Singh, *Beginning of Artificial Intelligence, End of Human Rights?*, LSE BLOG (July 16, 2020), https://blogs.lse.ac.uk/humanrights/2020/07/16/beginning-of-artificial-intelligence-end-of-human-rights/.

231 *Put Artificial Intelligence to Work for You*, EURES (Mar. 22, 2024), https://eures.europa.eu/put-artificial-intelligence-work-you-2024-03-22_en.

232 *See e.g., Employment Differences Between US and Europe*, EURDEV (Sep 19, 2023), https://www.eurodev.com/blog/employment-differences-between-us-and-europe?utm_source=chatgpt.com.

General Strike, and the rise of the Communist Party USA in the 1920s—were swiftly suppressed.[233] The Red Scares of the 1920s and 1950s dismantled organized labor's capacity to push for systemic protections.[234] Without a history of large-scale worker uprisings shaping its policies, [235] the U.S. regulatory framework has largely prioritized corporate freedom over labor safeguards.[236]

The US approach to AI regulation is consistent with its long-standing management style. Since the early 20th century, U.S. management philosophies have focused on productivity, innovation, and adaptability, sidelining uniform safeguards and centralized oversight.[237] Frederick Winslow Taylor's *The Principles of Scientific Management* treated workers as

---

233 *See e.g.,* Jill Lepore, *Eugene V. Debs and the Endurance of Socialism*, THE NEW YORKER (Feb. 11, 2019), https://www.newyorker.com/magazine/2019/02/18/eugene-v-debs-and-the-endurance-of-socialism; Darrin Hoop, *Seattle: the 1919 General Strike*, INTERNATIONAL SOCIALIST REVIEW, https://isreview.org/issue/84/seattle-1919-general-strike/index.html. (last visited Nov 15, 2024); Norman Markowitz, *The Communist Party in the 1920: The First Decade of Struggle*, PEOPLE'S WORLD (April 24, 2019), https://www.peoplesworld.org/article/the-communist-party-in-the-1920s-the-first-decade-of-struggle/. See also *Communist Party USA History and Geography*, UNIVERSITY OF WASHINGTON https://depts.washington.edu/moves/CP_intro.shtml (last visited Nov. 15, 2024).

234 *See Red Scare*, HISTORY.COM (updated April 21, 2023) https://www.history.com/topics/cold-war/red-scare.

235 *See generally* SEYMOUR MARTIN LIPSET & GARY MARKS, IT DIDN'T HAPPEN HERE: WHY SOCIALISM FAILED IN THE UNITED STATES (W.W. Norton & Co. 2000) (examining the aspects of American society that impeded the development of socialist and communist movements); IRVING HOWE & LEWIS A. COSER, THE AMERICAN COMMUNIST PARTY: A CRITICAL HISTORY, 1919-1957 (Beacon Press 1957) (offering a detailed analysis of the Communist Party USA, exploring its internal dynamics, external challenges, and the reasons behind its limited impact on American politics).

236 Organized labor in the U.S. achieved some victories, especially during the New Deal era, but these were tempered by the dominance of business interests and the individualistic "American Dream." Unlike the EU, which developed worker protections from a history of labor oppression, the U.S. lacked a comparable catalyst. Instead, industries relied on government support to curb union power, as seen in the Taft-Hartley Act of 1947, which limited strikes and political activities. It has always been the invisible hand of competition—not collective struggle—that has always shaped Labor. However, recent developments suggest a growing awareness of the risks posed by unregulated AI in the workplace. The Federal Trade Commission (FTC) has begun investigating cases of algorithmic bias and unfair practices, while state-level privacy laws, such as the California Consumer Privacy Act (CCPA), are starting to impose stricter rules on data use. Although these steps fall short of the EU's proactive regulatory model, they signal a shift toward greater accountability for AI-driven workplace practices.

237 E.g., Some of the notable management theories include Frederick Winslow Taylor's Scientific Management, which emphasizes efficiency and productivity through systematic observation and measurement of work processes; Henri Fayol's Administrative Management Theory, which focuses on the managerial practices necessary for organizational efficiency; the Behavioral Management theory which emphasizes the importance of human behavior, needs, and attitudes within the workplaces; and the post-World World II, modern management theories which combines mathematical principles with sociology to create holistic approaches to management.

interchangeable machine parts,[238] with efficiency outweighing individual autonomy.[239] Later, Mary Parker Follett brought in ideas like collaboration and power-sharing.[240] But even her approach prioritized boosting business performance.[241] This pragmatic mindset continues to define the US AI regulation, favoring flexible, business-led frameworks over rigid, top-down controls. The U.S. Algorithmic Accountability Act proposed in 2003 reflected this philosophy, but even that modest bill died on the vine.[242] Nonetheless, at least one state stepped into this terrain. New York City's Automated Employment Decision Tools (AEDT) law requires bias audits for AI hiring tools and mandates notifying candidates of AI use.[243] But businesses still call the shots—they pick the auditors, define the audit methods, and decide how to handle the

238 *See Taylorish*, MUNICH BUSINESS SCHOOL, https://www.munich-business-school.de/en/l/business-studies-dictionary/taylorism (last visited Nov. 16, 2024) (reducing human worker to a kind of machine and the separation of thought and action)

239 *See* Hannah Taylor, *What is Taylorism & Why You Should Think Beyond It*, RUN, https://www.runn.io/blog/what-is-taylorism.

240 *See Mary Parker Follett on Community, Creativity, and Control*, MICHELE ZANINI (Nov 18, 2020), https://www.michelezanini.com/mary-parker-follett-the-first-prophet-of-management//

241 *See* Louise Delaney, *Mary Parker Follett: Management Thought Leader of the Early 1900s*, MONARCH, https://umonarch.ch/2023/03/16/mary-parker-follett-management-thought-leader-of-the-early-1900s/ (last visited Nov 16, 2024) She emphasized a holistic management using an integrative approach where employee needs and business objectives are harmonized. Believing in concepts like "integration" and "co-active power," she proposed that employee well-being wasn't an end in itself but a means to create a more effective and cooperative workplace. After employees felt valued and motivated, their contributions to businesses improved.

242 *See* Jakob Mökander et al., *The U.S. Algorithmic Accountability Act of 2022 vs. The EU Artificial Intelligence Act: What Can They Learn from Each Other?*, 32 MINDS & MACHINES 751 (2022), https://link.springer.com/article/10.1007/s11023-022-09612-y (arguing that the AAA mandates that companies conduct impact assessments for automated decision systems to enhance transparency and accountability but also provide them with significant discretion). *See* Joshua New, *How to Fix the Algorithmic Accountability Act, Center for Data Innovation*, CENTER FOR DATA INNOVATION (Sep 23, 2019), https://datainnovation.org/2019/09/how-to-fix-the-algorithmic-accountability-act/ (noting that the bill "misses the mark" by holding algorithms to different standards than humans and targeting only large firms); *See also* Maneesha Mithal, Gabriella Monahova, and Andrew Stivers, *The Algorithmic Accountability Act: Potential Coverage Gaps in the Healthcare Sector*, AMERICAN BAR ASSOCIATION (August 2022), https://www.americanbar.org/content/dam/aba/publications/antitrust/magazine/2022/august/algorithmic-accountability-act.pdf.

243 *See Automated Employment Decision Tools*, NYC CONSUMER AND WORKER PROTECTION, https://www.nyc.gov/site/dca/about/automated-employment-decision-tools.page (last visited Nov 15, 2024).

results.[244] Compliance, in the end, is their choice.[245] Similarly, Illinois' Artificial Intelligence Video Interview Act demands consent before AI is used to analyze video interviews[246] but leaves terms like "artificial intelligence" vague, giving companies an easy out if they want to sidestep the law.[247]

California's Consumer Privacy Act (CCPA) gives employees the right to opt out of AI-driven decisions,[248] but these rights may be of little practical benefit. Opting out can make workers look like they are resisting technology, potentially putting their jobs or promotions at risk.[249] Many avoid exercising these rights altogether, making the protections largely symbolic. Taken together, the failure of the Algorithmic Accountability Act, the modest protection afforded by New York's AEDT law, Illinois' AI Video Interview Act, and the limited protections in the CCPA embody the United States's "move fast and break things" ethic.[250] For over a century, the focus has been on scaling up, innovating, and sticking to pragmatic governance, rather than imposing top-down controls.

*

244 *See* Lara Groves, Jacob Metcalf, Alayna Kennedy, Briana Vecchione & Andrew Strait, *Auditing Work: Exploring the New York City Algorithmic Bias Audit Regime*, in PROCEEDINGS OF THE 2024 ACM CONFERENCE ON FAIRNESS, ACCOUNTABILITY, AND TRANSPARENCY (FACCT '24) 1107 (June 3–6, 2024), https://doi.org/10.1145/3630106.3658959 (arguing that the law lacks clear definitions of what constitutes an independent auditor, suggesting that companies have the discretion to determine the extent and methods of the audits). *See also e.g.,* Ifeoma Ajunwa, *An Auditing Imperative for Automated Hiring*, 34 HARV. J.L. & TECH. 621 (2021).

245 Although annual audits are mandated, companies are not required to publicly disclose full findings. *See Frequently Asked Questions About Local Law 144 of 2021: Automated Employment Decision Tools*, NYC DEP'T OF CONSUMER & WORKER PROT. (June 2023), https://www.nyc.gov/assets/dca/downloads/pdf/about/DCWP-AEDT-FAQ.pdf (employers are only required to share the *summary* of the most recent bias audit, not the full report).

246 *See* Daniel Walt et al., *Illinois Employers Must Comply with Artificial Intelligence Video Interview Act*, SHRM (Sep 4, 2019), https://www.shrm.org/topics-tools/employment-law-compliance/illinois-employers-must-comply-artificial-intelligence-video-interview-act.

247 *See* Aaron J. Burstein, *Employers Beware: The Illinois Artificial Intelligence Video Interview Act is Now in Effect*, KELLEY DRYE (Jan 15, 2020), https://www.kelleydrye.com/viewpoints/blogs/ad-law-access/employers-beware-the-illinois-artificial-intelligence-video-interview-act-is-now-in-effect.

248 *See California Consumer Privacy Act*, ROB BONTA ATTORNEY GENERAL (updated March 13, 2024), https://oag.ca.gov/privacy/ccpa.

249 *See Fact Sheet: Retaliation Based on Exercise of Workplace Rights Is Unlawful*, U.S. EQUAL EMPLOYMENT OPPORTUNITY COMMISSION, https://www.eeoc.gov/fact-sheet-retaliation-based-exercise-workplace-rights-unlawful (last visited Nov. 15, 2024) (acknowledging that many workers are deterred from asserting workplace rights due to fear of retaliation).

250 *See generally* JONATHAN TAPLIN, MOVE FAST AND BREAK THINGS: HOW FACEBOOK, GOOGLE, AND AMAZON CORNERED CULTURE AND UNDERMINED DEMOCRACY (Illus. ed. Little, Brown & Co. 2017).

Once again, the foregoing discussion illustrates that EU AI regulation is less a reflection of a rights-based normative vision and more a pragmatic institutional response aimed at preserving systemic stability and mitigating technological risks within a complex regulatory landscape. Rooted in a historical awareness of labor unrest and systemic exploitation, EU labor regulations are designed not merely to protect individual rights but to preempt instability and maintain social cohesion by ensuring that AI does not become a tool of unchecked employer control. This approach prioritizes risk mitigation over ideological commitments. Unlike the U.S., which emphasizes corporate freedom and innovation, the EU prioritizes regulatory oversight to prevent AI from exacerbating inequalities that might foster discontent and destabilization.

## E. AI-Generated Misinformation and Disinformation

Foreign influence campaigns, viral conspiracy theories, and deepfakes designed to reduce public trust, sow discord, or inflict emotional distress are now a fact of modern life. Although online misinformation and disinformation have emerged as pressing threats to civic order and democratic stability on both sides of the Atlantic, policymakers in the in the EU and the U.S. have responded with markedly different regulatory approaches. The regulation of AI-generated misinformation and disinformation is yet another key area where, far from being anchored in a rights-based normative philosophy, EU rules constitute a pragmatic framework shaped by Europe's historical and cultural imperatives—one designed to balance technological advancement with stability and structured risk management. The EU addresses AI-driven misinformation through a unified regulatory framework that emphasizes platform accountability, privacy, and transparency.[251] Europe's approach prioritizes preventive risk management over individual free speech, whereas the American approach, rooted in the First Amendment, prioritizes free expression even at the expense of a more chaotic (and arguably toxic) information environment.[252]

The EU's approach might be better understood when contextualized in its history of disinformation being used to destabilize societies and disrupt political order. Throughout its history, false information has been a powerful tool for manipulating public opinion, undermining trust in institutions, and inciting unrest. During the French Revolution, the spread of rumors like the "Great Fear"—which falsely claimed that peasants were under attack by aristocrats—sparked widespread panic and violent uprisings, ultimately leading to attacks on estates, the looting of grain supplies, and the ultimate breakdown of feudal structures. [253] In the 20th century, Nazi Germany used propaganda to consolidate power.[254] Joseph Goebbels, Hitler's propaganda minister, harnessed the media of the time—radio, print, and film—to push the regime's message.[255] Films like *The Eternal Jew* spread anti-Semitic stereotypes,[256] while books such as *The Myth of the Twentieth Century* by Alfred Rosenberg provided a pseudo-historical framework to justify discriminatory laws like the Nuremberg Race Laws.[257]

---

[251] *See generally* Andreu Casero-Ripollés, Jorge Tuñón & Luis Bouza-García, *The European Approach to Online Disinformation: Geopolitical and Regulatory Dissonance*, 10 HUMANITY & SOC. SCI. COMMC'NS, ART. 657 (2023).

[252] *See* Will Rasenberger, *Does the First Amendment Protect AI-Generated Speech?,* THE REGULATORY REVIEW (Mar 19, 2024), https://www.theregreview.org/2024/03/19/rasenberger-does-the-first-amendment-protect-ai-generated-speech/. *See e.g., Broadcast News Distortion,* FED. COMMC'NS COMM'N, https://www.fcc.gov/broadcast-news-distortion (last visited Nov. 16, 2024) (FCC's role involves oversight of content but is limited in directly regulating misinformation due to First Amendment protections); *See Advertising and Disclaimers*, FED. ELECTION COMM'N, https://www.fec.gov/help-candidates-and-committees/advertising-and-disclaimers/ (last visited Nov. 16, 2024) (the FEC monitors the transparency and accuracy of disclosures in political advertisements and spending but has limited power to address misinformation directly.)

The European Union has adopted an increasingly interventionist approach to online content governance to counter disinformation. Early efforts relied on co-regulation, such as the 2018 Code of Practice on Disinformation, in which major platforms pledged voluntary compliance.[258] When the Code was introduced, EU officials signaled that without progress, binding legislation would follow and it was not long before they judged that the Code was not sufficient in light of persistent foreign propaganda and COVID-19 falsehoods.[259]

The European Union's Digital Services Act (DSA), enacted in 2022, represents a sweeping regulatory intervention targeting online intermediaries, including social media networks and

---

[253] *See The Great Fear*, ENCYCLOPAEDIA BRITANNICA, https://www.britannica.com/event/Great-Fear (last visited Jan. 6, 2025); *See also The Great Fear*, WORLD HISTORY ENCYCLOPEDIA, https://www.worldhistory.org/Great_Fear (last visited Jan. 6, 2025).

[254] *See e.g., Nazi Propaganda and Censorship*, HOLOCAUST ENCYCLOPEDIA https://encyclopedia.ushmm.org/content/en/article/nazi-propaganda-and-censorship

(last visited Jan.7, 2025), although it must be conceded that other factors such as national discontent plays a role too, as causes of historical events are far from so linear and clear.

[255] *Id.*

[256] *Der Ewige Jude*, HOLOCAUST ENCYCLOPEDIA https://encyclopedia.ushmm.org/content/en/article/der-ewige-jude (last visited Jan.7, 2025)

[257] *See* Carmen Lea Dege & Tae-Yeoun Keum, *Editors' Introduction: Political Myth in the Twentieth Century*, 46 HISTORY OF EUROPEAN IDEAS 1199 (2023), https://doi.org/10.1080/01916599.2023.2198541.

[258] Mauro Fragale & Valentina Grilli, *Deepfake, Deep Trouble: The European AI Act and the Fight Against AI-Generated Misinformation,* COLUM. J. EUR. L. PRELIMINARY REFERENCE BLOG (Nov. 11, 2024), https://cjel.law.columbia.edu/preliminary-reference/2024/deepfake-deep-trouble-the-european-ai-act-and-the-fight-against-ai-generated-misinformation/.

[259] *Id.*

search engines.[260] At its core, the DSA aims to establish a "safe and accountable online environment," underscoring the EU's commitment to curbing the spread of disinformation.[261] The law imposes a series of obligations on major platforms—those exceeding 45 million users in the EU—including risk assessment mandates, independent audits, and stringent content moderation protocols. Rather than dictating specific speech restrictions, the DSA introduces procedural mechanisms designed to mitigate "systemic risks," thereby aligning regulatory enforcement with broader societal concerns rather than rigid doctrinal limits on speech.[262] One of the key mechanisms is that the DSA requires platforms to collaborate with trusted flaggers and fact-checkers.[263] The DSA also mandates transparency in algorithms and the disclosure of content moderation practices, making sure users understand how information is prioritized and filtered.[264] This procedural approach to combating online misinformation echoes prior national initiatives, such as Germany's *Netzwerkdurchsetzungsgesetz* (NetzDG), but supersedes them in favor of a harmonized EU-wide standard.[265] It is notable here that, given the dominance of U.S.-based tech companies in the European market, the DSA's provisions effectively extend European regulatory influence over online speech well beyond the EU's borders.

In addition to the DSA, the recently enacted EU AI Act also has implications for online misinformation. Under the Act's tiered risk paradigm, AI applications used for mass manipulation, including those facilitating disinformation campaigns, fall into the high-risk category, thereby triggering enhanced compliance requirements.[266] The AI Act mandates transparency obligations for developers of AI systems capable of generating realistic text, speech, or video content.[267] Under the Act, AI-generated materials—such as fabricated political speeches or synthetic news footage—must be clearly labeled, ensuring viewers are

260 *See The Digital Services Act Package*, EUROPEAN COMMISSION, https://digital-strategy.ec.europa.eu/en/policies/digital-services-act-package (last visited Feb 10, 2025)

261 *See The Digital Services Act,* EUROPEAN COMMISSION https://commission.europa.eu/strategy-and-policy/priorities-2019-2024/europe-fit-digital-age/digital-services-act_en (last visited Feb.11, 2025)

262 *Id.*

263 *Id.*

264 *Id.*

265 *See generally* Patrick Zurth, *The German NetzDG as Role Model or Cautionary Tale? Implications for the Debate on Social Media Liability*, 31 FORDHAM INTELL. PROP. MEDIA & ENT. L.J. 1084 (2021).

266 *See Risk-Classifications According to the EU AI Act*, TRAIL ML BLOG, https://www.trail-ml.com/blog/eu-ai-act-how-risk-is-classified (last visited Nov 16, 2024)

267 *See EU AI Act: First Regulation on Artificial Intelligence*, EUROPEAN PARLIAMENT (June 18, 2024), https://www.europarl.europa.eu/topics/en/article/20230601STO93804/eu-ai-act-first-regulation-on-artificial-intelligence.

made aware of the content's artificial origins.[268] This legislative effort underscores the EU's broader regulatory philosophy: rather than banning high-risk AI outright, the Act seeks to preempt foreseeable harms through mandated transparency and accountability mechanisms.

These hard-law requirements are supplemented by soft-law in the form of policy initiatives such as the European Democracy Action Plan (EDAP)[269] and voluntary self-regulatory frameworks such as the EU Code of Practice on Disinformation[270] and the European Digital Media Observatory.[271] While the EDAP itself is not legally binding, it has led to legislative proposals, such as the Regulation on Transparency and Targeting of Political Advertising (adopted in 2023), updates to the Code of Practice on Disinformation and strengthened rules under the DSA and Media Freedom Act.[272] Companies such as Google, Facebook, Twitter, and Microsoft have signed the EU Code of Practice on Disinformation, pledging to label AI-generated content, remove harmful disinformation, demonetize its sources, and increase transparency, particularly during elections.[273] Although the Code began as a classic soft-law instrument in 2018,[274] failure to meet its commitments could contribute to enforcement actions under the DSA, which is hard law. We also note the importance the EU attaches to cross-border collaboration on combating misinformation. The European External Action Service (EEAS) StratCom Task Forces actively track and counter state-sponsored disinformation, especially from Russia.[275]

---

268 *Id.*

269 *See Strengthened EU Code of Practice on Disinformation*, EUROPEAN COMMISSION, https://commission.europa.eu/strategy-and-policy/priorities-2019-2024/new-push-european-democracy/protecting-democracy/strengthened-eu-code-practice-disinformation_en (last visited Nov 16, 2024)

270 *See e.g.,* Giorgio Borz et al., *The EU Soft Regulation of Digital Campaigning: Regulatory Effectiveness Through Platform Compliance to the Code of Practice on Disinformation*, 45 Pol'y Stud. 709 (2024).

271 *European Digital Media Observatory (EDMO)*, EUROPEAN COMMISION, https://digital-strategy.ec.europa.eu/en/policies/european-digital-media-observatory (last visited Nov. 16, 2024) (a collaborative platform that brings together fact-checkers, media literacy experts, and academic researchers to understand and address disinformation).

272 See EU Introduces New Rules on Transparency and Targeting of Political Advertising, EUROPEAN COUNCIL (March 11, 2024), https://www.consilium.europa.eu/en/press/press-releases/2024/03/11/eu-introduces-new-rules-on-transparency-and-targeting-of-political-advertising/; *See also Protecting Democracy*, EUROPEAN COMMISSION https://commission.europa.eu/strategy-and-policy/priorities-2019-2024/new-push-european-democracy/protecting-democracy_en (last visited Feb. 10, 2025)

273 *See The 2022 Code of Practice on Disinformation*, EU DIGITAL STRATEGY, https://digital-strategy.ec.europa.eu/en/policies/code-practice-disinformation (last visited Jan. 6, 2025).

274 *Id.*

275 *See Tackling Disinformation: Information Work of the EEAS Strategic Communication Division and Its Task Forces*, EUROPEAN EXTERNAL ACTION SERV. (Oct 12, 2021), https://www.eeas.europa.eu/countering-

During moments of crisis, EU authorities have demonstrated a willingness to intervene directly in the digital information space. In response to Russia's invasion of Ukraine, the EU took the extraordinary step of banning state-controlled media outlets RT and Sputnik in 2022, citing their role in disseminating war propaganda.[276] At the national level, European states have reinforced these efforts through domestic legislation. France, for instance, enacted a law targeting disinformation in electoral contexts,[277] while Germany's NetzDG statute mandates swift removal of unlawful and misleading content.[278] These interventions illustrate a broader regulatory philosophy: a proactive, state-driven approach to structuring the online information ecosystem in order to fortify democratic institutions and maintain social cohesion. Rather than relying on market self-regulation, European regulators embrace a model of legal constraint designed to preempt the harmful spread of falsehoods. Crucially, these measures are framed not as censorship but as necessary safeguards against concrete threats such as hate speech, election interference, and public health misinformation.[279] Free expression remains a core European value, yet it is balanced against competing societal imperatives; the dominant perspective holds that extreme or demonstrably false speech may be lawfully restricted to protect the public.[280] As a result, Europe has developed a stringent regulatory framework that imposes affirmative obligations on online platforms, treating them as custodians of the digital public sphere responsible for mitigating the risks of viral disinformation. In the EU, individual rights to free expression often yield to the broader public interest in maintaining political and social stability and managing risks related to AI misinformation.

*

Whereas the EU has implemented a range of regulatory measures to address the proliferation of online disinformation, the United States, by contrast, has no comparable statutory framework. This is no accident: any such law would face formidable constitutional disinformation/tackling-disinformation-information-work-eeas-strategic-communication-division-and-its-task-forces_und_en.

[276] *See EU Officially Boots Russia's RT, Sputnik Outlets*, POLITICO https://www.politico.eu/article/russia-rt-sputnik-illegal-europe/ (last visited Jan.7, 2025).

[277] *See* Zachary Young, *French Parliament Passes Law Against "Fake News"*, POLITICO (July 4, 2018), https://www.politico.eu/article/french-parliament-passes-law-against-fake-news/.

[278] *See* Diana Lee, *Germany's NetzDG and the Treat to Online Free Speech*, YALE LAW SCHOOL (Oct. 10, 2017), https://law.yale.edu/mfia/case-disclosed/germanys-netzdg-and-threat-online-free-speech.

[279] *See e.g., Germany: Network Enforcement Act Amended to Better Fight Online Hate Speech*, LIBRARY OF CONGRESS, https://www.loc.gov/item/global-legal-monitor/2021-07-06/germany-network-enforcement-act-amended-to-better-fight-online-hate-speech/ (last visited Feb. 10, 2025).

[280] *Id.*

challenges.[281] The First Amendment, a cornerstone of American legal tradition, imposes stringent limitations on government action restricting speech.[282] Under well-established Supreme Court precedent, no broad exception exists for falsehoods; even demonstrably false or misleading statements generally remain protected unless they constitute a separate, legally cognizable harm—such as fraud, defamation, or imminent threats of violence.[283] The current Supreme Court's First Amendment maximalism is such that almost any direct government regulation of online content would be held unconstitutional.[284] As has been discussed extensively in legal scholarship,[285] regulating misinformation would likely qualify as a content-based restriction, triggering strict scrutiny.[286] The government must prove that a law serves a compelling state interest and is narrowly tailored—standards misinformation regulations rarely meet.[287] On top of that, federal attempts risk a chilling effect, where platforms and individuals over-censor legitimate speech to avoid liability.[288] The *United States v. Alvarez* decision is one of the most important cases to consider in this scenario—it ruled that even false speech is protected under the First Amendment unless it directly causes harm, like defamation or fraud.[289] Without clear evidence of harm, regulating AI-generated misinformation remains a constitutional minefield.[290]

---

281 *See Why Regulating AI Will Be Difficult or Even Impossible*, TRAILS https://www.trails.umd.edu/news/why-regulating-ai-will-be-difficult-or-even-impossible (last visited Nov. 16, 2024).

282 U.S. CONST. AMEND. I.

283 U.S. v. Alvarez, 567 U.S. 709 (2012); U.S. v. Stevens, 559 U.S. 460 (2010); Brandenburg v. Ohio, 395 U.S. 444 (1969).

284 The current Supreme Court interprets the First Amendment in a way that strongly protects free speech, even in cases where the government wants to regulate harmful online content. If the government tries to pass laws directly controlling what can or cannot be posted on the internet, the Court will likely strike them down as unconstitutional.

285 *See e.g.,* Ari E. Waldman, *The Marketplace of Fake News,* 20 U. PA. J. CONST. L. 845 (2018); Randy J.Kozel, *Content Under Pressure*, 100 WASH. U. L. REV. 59 (2022).

286 *See False Speech and the First Amendment: Constitutional Limits on Regulating Misinformation*, CONGRESSIONAL RESEARCH SERVICE (August 1, 2022), https://crsreports.congress.gov/product/pdf/IF/IF12180.

287 *Id.*

288 *Id.*

289 Alvarez, 567 U.S. 709 (2012) ("given our historical skepticism of permitting the government to police the line between truth and falsity, and between valuable speech and drivel, we presumptively protect all speech, including false statements, in order that clearly protected speech may flower in the shelter of the First Amendment.").

290 There are, however, proposed federal bills such as the Malicious Deep Fake Prohibition Act and Online False Information Accountability Act. The former criminalizes creating and distributing deepfakes intended to deceive, defraud, or cause harm. It applies only to specific cases like election fraud, defamation, or financial fraud, leaving broader issues of deepfake-driven misinformation, such as public health disinformation or targeted personal

In view of these limitations, rather than enacting outright prohibitions on false content or punishing those who propagate conspiracy theories—both of which would violate First Amendment principles—lawmakers have pursued disclosure-based measures, such as mandating transparency in online political advertising and social media algorithms.[291] For example, instead of outlawing deepfakes, the previous Congress considered, but did not enact, the Deep Fakes Accountability Act to encourage watermarks or disclosures on AI-altered media.[292] Agencies like the Federal Election Commission have updated guidelines to extend campaign advertisement disclosure rules to personal websites and advertising platforms.[293] The U.S. government has also been advised to fund digital literacy initiatives to help citizens better recognize false information online.[294] These approaches reflect a foundational tenet of American free speech jurisprudence: that the appropriate response to falsehoods is not suppression but counterspeech, factual rebuttal, and the free exchange of ideas.

At the state level, responses are narrowly focused and highly specific, with most regulations aimed at elections or the commercial misuse of likenesses.[295] For example, California's *Defending Democracy from Deep Fake Deception Act* requires platforms to label or remove deceptive, digitally altered election content during election periods, allowing candidates and officials to seek injunctive relief.[296] Similarly, Texas' *Deep Fake Law* criminalizes the use of deep fake videos to mislead or harm voters during elections but doesn't extend to misinformation outside that scope.[297]

---

disputes, unaddressed. The latter holds digital platforms accountable for disseminating AI-generated misinformation. *See* S. 3805, 115th Cong. (2018); H.R. 3230, 116th Cong. (2019).

291 *See e.g.,* Max I. Fiest, *Why a Data Disclosure Law Is (Likely) Unconstitutional*, 43 COLUM. J.L. & ARTS 517, 524-526 (2020).

292 H.R. 5586, 118th Cong. (2023).

293 *Commission Adopts Final Rule on Internet Communications Disclaimers and the Definition of Public Communication*, FEDERAL ELECTION COMMISSION (Dec. 19, 2022), https://www.fec.gov/updates/commission-adopts-final-rule-internet-communications-disclaimers-and-definition-public-communication/.

294 *See* Jon Bateman and Dean Jackson, *Countering Disinformation Effectively: An Evidence-Based Policy Guide*, CARNEGIE ENDOWMENT FOR INTERNATIONAL PEACE (Jan. 31, 2024), https://carnegieendowment.org/research/2024/01/countering-disinformation-effectively-an-evidence-based-policy-guide?lang=en.

295 *See* Travis Yuille, *Analysis: Elections and Obscenity Will Continue Driving AI Laws*, BLOOMBERG LAW (Nov 11, 2024), https://news.bloomberglaw.com/bloomberg-law-analysis/analysis-elections-and-obscenity-will-continue-driving-ai-laws.

296 *See generally,* A.B. 2655, 2024 Leg., Reg. Sess. (Cal. 2024).

297 *See* Tori Guidry, *Texas Did It First: Texas Was the First to Enact State Legislation on the Use of Deep Fakes in Elections*, NAT'L L. REV. (May 28, 2024), https://natlawreview.com/article/texas-did-it-first-texas-was-first-enact-state-legislation-use-deep-fakes-elections.

Beyond these election-focused laws, some states have broader measures targeting specific harms from misinformation. New York's *Civil Rights Law* prohibits the unauthorized use of a person's name, portrait, or picture for advertising or trade purposes without consent, but it doesn't cover purely expressive or non-commercial uses.[298] Similarly, Washington's *Impersonation and Defamation Statute* criminalizes the use of someone else's likeness for malicious purposes, such as defamation or impersonation in campaign materials, though its enforcement has been rare and largely symbolic.[299]

Publicity laws, such as Illinois' *Right of Publicity Act*, address unauthorized commercial use of individuals' likenesses or personal attributes for profit but fail to cover non-commercial uses or broader privacy concerns.[300] States like Massachusetts and New York have similar laws, but their scope is likewise limited.[301] Tennessee expanded and renamed its 1984 right of publicity law to account for the threat that generative AI was thought to pose to the music industry with the "Ensuring Likeness, Voice, and Image Security Act of 2024" or "ELVIS" Act.[302] The law adds voice to the personal attributes protected along with the traditional "name, image, and likeness."[303] Significantly, the revised law also creates liability for publishing, performing, distributing, transmitting, or otherwise making available an individual's voice, likeness, or the means to create that voice or likeness, with knowledge that the use was unauthorized.[304] The breadth of this law is offset by a First Amendment savings clause that provides "To the extent such use is protected by the First Amendment to the United States Constitution, it is deemed a fair use and not a violation of an individual's right…" in certain contexts.[305] This drafting leaves those subject to the law with little option but to become experts in the First Amendment to understand what they may and may not do with generative AI impersonation tools. But

298 Judith B. Bass, *New York's New Right of Publicity Law: Protecting Performers and Producers*, NYSBA (March 17, 2021), https://nysba.org/new-yorks-new-right-of-publicity-law-protecting-performers-and-producers.

299 *See* John Thayer, *Defamation or Impersonation? Working Towards a Legislative Remedy for Deepfake Election Misinformation*, 66 WM. & MARY L. REV. 251, 263 (2024),

300 *See* 765 ILCS 1075/10. *See also Illinois Right of Publicity Act*, HINSHAW & CULBERTSON LLP, https://www.hinshawlaw.com/services-illinois-right-of-publicity-act.html. (last visited Nov. 17, 2024)

301 *See* RIGHT OF PUBLICITY, https://rightofpublicity.com/statutes/massachusetts (last visited Nov. 17, 2024); Cal. Civ. Code §§ 3344–3346 (West 2024); *New York Right of Publicity Law*, DIGITAL MEDIA LAW PROJECT (September 9, 2024), https://www.dmlp.org/legal-guide/new-york-right-publicity-law. *See generally* STATE LAW: RIGHT OF PUBLICITY, DIGITAL MEDIA LAW PROJECT (September 9, 2024), https://www.dmlp.org/legal-guide/state-law-right-publicity (for a list of state laws in relation to right of publicity).

302 Tenn. Code Ann. § 47-25-1101 to 1108, as amended by 2024 Tenn. Pub. Acts 588.

303 Tenn. Code Ann. § 47-25-1105.

304 *Id.*

305 Tenn. Code Ann. § 47-25-1107.

ELVIS aside, significant gaps still remain in addressing misinformation and privacy issues due to the First Amendment.

Any state effort to regulate online speech must also grapple with the broad immunity from state law implied by Section 230 of the Communications Decency Act.[306] Enacted in 1996, Section 230 shields online intermediaries from liability for third-party content, affording platforms wide latitude to curate, host, or remove user-generated speech at their discretion. Dubbed "the twenty-six words that created the internet,"[307] this statute is credited with catalyzing the rise of social media by eliminating platform liability for most user content. Consequently, U.S. tech companies have no legal duty to remove user generated misinformation—unless it happens violates a federal intellectual property right or some other narrow exception—nor do they face liability for failing to act.[308] Content moderation by platforms such as X, Facebook, and YouTube are voluntary and seem increasing fragile.[309] Attempts by government actors to dictate platform content moderation policies have encountered constitutional roadblocks. For instance, recent state laws in Florida and Texas, which sought to prevent social media platforms from removing content based on viewpoint, were promptly challenged in court as violations of private companies' First Amendment rights.[310]

The First Amendment, positioned at the forefront of the Bill of Rights, embodies a foundational principle of American constitutionalism: the protection of expressive freedom as a check on power. Across history, it has served as both a shield and a catalyst, invoked in seminal cases to reaffirm the right to resist orthodoxy and contest authority. In *Brandenburg v. Ohio*,[311] for instance, the Supreme Court ruled that even inflammatory speech advocating illegal action is protected unless it is likely to incite imminent lawless action;[312] in *Texas v. Johnson*,[313] the government cannot prohibit expression simply because it is offensive or provocative. Such a commitment to free speech has fostered a culture of innovation, open discourse, and dissent.

---

[306] 47 U.S.C. § 230(e)(3)

[307] *See generally* JEFF KOSSEFF, THE TWENTY-SIX WORDS THAT CREATED THE INTERNET (Cornell Univ. Press 2019).

[308] Hepp v. Facebook, 14 F.4th 204 (3d Cir. 2021).

[309] *See e.g., Community Standards*, META, https://transparency.meta.com/policies/community-standards/?source=https%3A%2F%2Fwww.facebook.com%2Fcommunitystandards (last visited Feb. 10, 2025)

[310] *See* Moody v. NetChoice, LLC, 603 U.S. 707 (2024); NetChoice, L.L.C v. Paxton, 121 F.4th 494 (5th Cir. 2024).

[311] 395 U.S. 444 (1969).

[312] *Id.*

[313] 491 U.S. 397 (1989).

As it becomes an integral part of the country's identity, and part of what makes America American, it's no surprise that, when faced with new challenges like AI-driven misinformation, the U.S. instinctively turns to the First Amendment, adhering to its principles rather than adopting the centralized, stability-driven regulatory measures favored by the EU.

The regulation of AI-generated misinformation and disinformation reveals a striking transatlantic contrast—one that does not merely challenge the prevailing fundamental rights paradigm but reverses it entirely. In the United States, a rights-centric legal framework—albeit one defined by a particularly constrained conception of rights—substantially limits efforts to regulate online content, even in contexts where the risks are significant. Meanwhile, European Union policymakers have adopted a markedly different approach, implementing systematic regulatory mechanisms to curb the dissemination of harmful digital falsehoods.

## III. CONCLUSION AND IMPLICATIONS

This Article has challenged the prevailing view that European Union AI regulation represents a principled commitment to fundamental rights. Through comparative institutional analysis across five regulatory domains, we have shown that the EU's approach reflects a deeper logic of risk management and administrative control rather than moral universality. The implications of this reframing extend beyond academic debate.

First, it calls into question the assumption that the EU model should serve as a global template for AI governance. The EU's regulatory framework emerged from specific historical conditions—fragmented sovereignty, imperial competition, and the ongoing need to balance competing state interests. These conditions do not exist everywhere, and the regulatory solutions they produced may not translate to other contexts. Moreover, the EU has chosen stability and constraint ahead of growth and some versions of individual autonomy. Other jurisdictions should make their own assessment of these tradeoffs rather than assuming the applicability or supremacy of the European template.

Second, our analysis suggests that debates over AI governance should focus less on competing values writ large and more on the details of institutional design. In other words, academics and policy advocates would do well to focus less on grand narratives that reflect our priors and more on the messy details of how rules and regulations work in practice. Questioning whether a system aspires to protect "human dignity" or promote "innovation" in the abstract will rarely lead to informative answers, especially if they displace more focused evaluation.

Third, this institutional perspective reveals that regulatory legitimacy takes more than moral branding. Claims about rights, ethics, and human-centric design may serve important rhetorical functions, but they cannot substitute for accountable governance structures. The EU's comprehensive regulatory apparatus may be more coherent than the U.S.'s fragmented approach, but coherence is not the same as legitimacy.

Our analysis does not argue that the U.S. model is superior to the European one, nor does it dismiss the importance of protecting individual rights. Rather, it suggests that these protections must be grounded in political processes and institutional arrangements that can sustain them over time. Rights without out of step with politics are merely aspirations.

The global AI governance landscape will likely remain pluralistic, with different jurisdictions adopting approaches that reflect their own political structures and regulatory traditions. This diversity should be welcomed. For scholars and policymakers, this means moving beyond debates about which jurisdiction has the "best" values and focusing instead on how different institutional arrangements actually function in practice. It means asking hard questions about who benefits from particular regulatory choices and how those choices can be made more democratically responsive. And it means recognizing that the legitimacy of AI governance will ultimately depend not on the nobility of its stated purposes, but on the quality of its institutional design and political accountability.

The stakes of getting this right are considerable. AI systems are reshaping economic relationships, political processes, and social interactions in ways that will affect generations to come. The regulatory frameworks we construct today will determine whether these technologies serve broader human flourishing or concentrate power in the hands of a few. Making that choice well requires clear thinking about governance, not just appealing rhetoric about rights.